\documentclass[aps,prd, twocolumn]{revtex4}
\usepackage{amsmath,amssymb,amsfonts}
\usepackage{graphicx}
\usepackage{enumerate} 
\usepackage{colordvi} 
\usepackage{bm}

\newcommand{\bfA}{\mbox{\boldmath$A$}}
\newcommand{\bfv}{\mbox{\boldmath$v$}}

\newcommand{\bfx}{\mbox{\boldmath$x$}}
\newcommand{\bfj}{\mbox{\boldmath$j$}}
\newcommand{\bfk}{\mbox{\boldmath$k$}}
\newcommand{\bfp}{\mbox{\boldmath$p$}}
\newcommand{\bfq}{\mbox{\boldmath$q$}}
\newcommand{\bfr}{\mbox{\boldmath$r$}}
\newcommand{\bfs}{\mbox{\boldmath$s$}}
\newcommand{\bfu}{\mbox{\boldmath$u$}}
\newcommand{\bfz}{\mbox{\boldmath$z$}}

\newcommand{\sigmav}{\sigma_{\rm v}}
\newcommand{\deltas}{\delta^{\rm(S)}}
\newcommand{\Pdd}{P_{\delta\delta}}
\newcommand{\Pdv}{P_{\delta \theta}}
\newcommand{\Pvv}{P_{\theta\theta}}
\newcommand{\DFoG}{D_{\rm FoG}}
\newcommand{\PKaiser}{P_{\rm Kaiser}}
\newcommand{\Plin}{P_{\rm lin}}
\begin{document}
\title{Baryon Acoustic Oscillations in 2D: Modeling Redshift-space 
Power Spectrum from Perturbation Theory}
\vfill
\author{Atsushi Taruya$^{1,2}$, Takahiro Nishimichi$^{2,3}$, Shun Saito$^{3,4}$}
\bigskip
\address{$^1$Research Center for the Early Universe, School of Science, 
The University of Tokyo, Bunkyo-ku, Tokyo 113-0033, Japan}
\address{$^2$Institute for the Physics and Mathematics of the Universe, 
University of Tokyo, Kashiwa, Chiba 277-8568, Japan}
\address{$^3$Department of Physics, The University of Tokyo, 
Bunkyo-ku, 113-0033, Japan}
\address{$^4$Department of Astronomy, 561-A Campbell Hall, 
University of California Berkeley, CA 94720, USA}
\bigskip
\date{\today}
%
\begin{abstract}
We present an improved prescription for
matter power spectrum in redshift space taking a 
proper account of both the non-linear
gravitational clustering and redshift distortion, 
which are of particular importance for accurately modeling 
baryon acoustic oscillations (BAOs). Contrary to the
models of redshift distortion phenomenologically introduced
but frequently used in the literature, 
the new model includes the corrections arising from the non-linear coupling 
between the density and velocity fields associated with two competitive 
effects of redshift distortion, i.e., 
Kaiser and Finger-of-God effects. Based on the improved 
treatment of perturbation theory for gravitational clustering, 
we compare our model predictions with monopole and quadrupole power spectra 
of N-body simulations, and 
an excellent agreement is achieved over the scales of BAOs.  
Potential impacts on constraining  
dark energy and modified gravity from the redshift-space power spectrum  
are also investigated based on the Fisher-matrix formalism. 
We find that the existing phenomenological models of redshift distortion 
produce a systematic error on measurements of 
the angular diameter distance and Hubble parameter by $1\sim2\%$,  
and the growth rate parameter by $\sim5\%$, which would become 
non-negligible for future galaxy surveys. 
Correctly modeling redshift distortion is thus essential, 
and the new prescription of redshift-space power spectrum 
including the non-linear corrections can be 
used as an accurate theoretical template for anisotropic BAOs. 
\end{abstract}

\maketitle

\maketitle

\section{Introduction}
\label{sec:intro}

Galaxy redshift surveys via the spectroscopic measurements of 
individual galaxies provide a three-dimensional map of galaxy 
distribution, which includes valuable cosmological information on 
structure formation of the Universe. The observed galaxy distribution is, 
however, apparently distorted due to the peculiar velocity of galaxies 
that systematically affects the redshift determination of each galaxy. 
The anisotropy caused by peculiar velocities is referred to as 
the {\it redshift distortion}, which 
complicates the interpretation of the galaxy clustering data (e.g., 
\cite{Hamilton:1997zq,Peebles:1980}).

Nevertheless, redshift distortion provides a unique way to measure 
the growth rate of structure formation, which has been previously 
used for determining the density parameters of the Universe 
(e.g., \cite{1992ApJ385L5H,Cole:1993kh}), 
and is now 
recognized with great interest as a powerful tool for testing 
gravity on cosmological scales 
(e.g., \cite{Linder:2007nu,Guzzo:2008ac,Yamamoto:2008gr,Song:2008qt,
Song:2010bk}). Redshift distortion 
also provides a helpful information on the dark-sector interactions 
\cite{Koyama:2009gd}, where the dark energy is dynamically coupled with 
dark matter (e.g., \cite{Zimdahl:2001ar,Farrar:2003uw}).  
Note that the distortion of the galaxy 
clustering pattern also arises from the apparent mismatch of the 
underlying cosmology when we convert the redshift and angular position 
of each galaxy to the comoving radial and transverse distances. This is 
known as Alcock-Paczynski effect \cite{Alcock_Paczynski:1979}, 
and with the baryon acoustic 
oscillations (BAOs) as a robust standard ruler, it can be utilized 
for a simultaneous measurement of the Hubble parameter $H(z)$ 
and angular diameter distance $D_A(z)$ of distant galaxies at 
redshift $z$ (e.g., \cite{Seo:2003pu,Blake:2003rh,Glazebrook:2005mb,
Shoji:2008xn,Padmanabhan:2008ag}).

In these respects, anisotropic clustering data from galaxy redshift surveys 
serve as a dual cosmological probe of the cosmic expansion and the gravity 
on cosmological scales, from which we can address properties of both the 
dark energy and modification of gravity responsible for the late-time 
cosmic acceleration. Although current data are not yet sensitive enough 
to separately measure $H(z)$, $D_A(z)$ and growth rate (see 
\cite{Reid:2009xm,Percival:2009xn,Kazin:2010nd,Yamamoto:2008gr} for 
current status), 
planned and ongoing galaxy redshift surveys aim at precisely 
measuring the anisotropic power spectrum and/or two-point 
correlation function in redshift space. Thus, 
the accurate theoretical modeling of anisotropic power spectrum 
is crucial and needs to be developed toward future observations.

The purpose of this paper is to address these issues based on the 
the analytical treatment of non-linear gravitational clustering. 
In the single-stream limit, cosmological 
evolution of the mass distribution consisting of the 
cold dark matter (CDM) and baryon is described by the coupled equations 
for irrotational and pressureless fluid \cite{Bernardeau:2001qr}. 
Recently, a detailed study on 
the standard treatments of perturbation theory has been made 
\cite{Jeong:2006xd,Jeong:2008rj,Nishimichi:2008ry}, 
and several improved treatments have been proposed 
\cite{Crocce:2005xy,Crocce:2005xz,Crocce:2007dt,Matsubara:2007wj,
      Matsubara:2008wx,McDonald:2006hf,Izumi:2007su,Taruya:2007xy,
      Taruya:2009ir,   
      Pietroni:2008jx,Matarrese:2007wc,Valageas:2003gm,Valageas:2006bi}, 
    showing that a percent-level accuracy can be achieved for the 
    predictions of 
power spectrum or two-point correlation function in real space. 
With a help of these, in this paper,  we will develop a model of 
redshift distortion, and compute the matter power spectrum in redshift space, 
with a particular attention to the BAOs. Also, we discuss the impact of 
model uncertainty of the redshift distortion on the 
acoustic-scale measurement of BAOs and the estimation of growth rate 
parameter.

This paper is organized as follows: In Sec.~\ref{sec:Pk_in_red}, 
we start with writing down the relation between real space and redshift 
space, and derive an exact expression for matter power spectrum in redshift 
space. We then consider the existing theoretical models of redshift 
distortion, and compare those with N-body simulations in 
Sec.~\ref{sec:existing_model}, showing that 
non-negligible discrepancy appears at the scales of BAOs.  
In Sec.~\ref{sec:improved_model}, non-linear corrections relevant to 
describe the small discrepancies  
are derived based on the exact expression of redshift-space power spectrum.  
The new model of redshift distortion including the corrections 
reproduces the N-body simulations quite well 
in both the monopole and quadrupole components of redshift-space power 
spectrum. In Sec.~\ref{sec:implication}, 
the relevance of this new model is discussed in details, 
especially for measurements of acoustic scales and growth rate parameters.  
The potential impact of the model of redshift distortion on future 
constraints on modified gravity 
and dark energy is also estimated based on Fisher matrix formalism. 
Finally, our important findings are summarized in Sec.~\ref{sec:conclusion}.

Throughout the paper, we assume a flat $\Lambda$CDM model and adopt 
the fiducial cosmological parameters based on the five-year WMAP results 
\cite{Komatsu:2008hk}: 
$\Omega_{\rm m}=0.279$, 
$\Omega_{\Lambda}=0.721$, $\Omega_{\rm b}/\Omega_{\rm m}=0.165$, $h=0.701$,
$n_s=0.96$, $\sigma_8=0.817$. In order to compare our analytic results with 
N-body simulations, the data are taken from Ref.~\cite{Taruya:2009ir}, in 
which $30$ independent N-body simulations of 
the $512^3$ particles and cubic boxes of side length $1h^{-1}$Gpc 
were carried out with initial conditions 
created by \verb|2LPT| code \cite{Crocce:2006ve} at $z_{\rm init}=31$, 
adopting the same cosmological parameters as mentioned above.

\section{Power spectrum in redshift space} 
\label{sec:Pk_in_red}

Let us first recall that the redshift distortion arises from 
the apparent mismatch of galaxy position between real and redshift spaces 
caused by the contamination of the peculiar velocities in the redshift 
measurement. For distant galaxies, the position in real space, 
$\bfr$, is mapped to the one in redshift space, $\bfs$, as
\begin{equation}
\bfs = \bfr + \frac{v_z(\bfr)}{a\,H(z)}\,\widehat{\bfz},
\label{eq:def_s-space}
\end{equation}
where the unit vector $\widehat{\bfz}$ indicates the line-of-sight 
direction, and quantity $v_z$ is the line-of-sight component of the 
velocity field, i.e, $v_z=\bfv\cdot\widehat{\bfz}$. The quantities $a$ 
and $H$ are the scale factor of the Universe and the Hubble parameter,  
respectively. Then, the 
density field in redshift space, $\deltas(\bfs)$, is related to 
the one in real space, $\delta(\bfr)$ through the relation 
$\{1+\deltas(\bfs)\}d^3\bfs=\{1+\delta(\bfr)\}d^3\bfr$, which leads to 
\begin{equation}
\deltas(\bfs)=\left|\frac{\partial\bfs}{\partial\bfr}\right|^{-1}
\left\{1+\delta(\bfr)\right\}-1.
\end{equation}
The Fourier transform of this is given by 
\begin{equation}
\deltas(\bfk)=\int d^3\bfr \left\{\delta(\bfr)-
\frac{\nabla_zv_z(\bfr)}{a\,H(z)}\right\}e^{i(k\mu\,v_z/H+\bfk\cdot\bfr)},
\end{equation}
where the quantity $\mu$ is the cosine of the angle between $\widehat{z}$ 
and $\bfk$. Here, we used the fact that the 
Jacobian $|\partial\bfs/\partial\bfr|$
is written as $1+\nabla_zv_z/(aH)$.

From this, the power spectrum of density in redshift space becomes
 \cite{Scoccimarro:2004tg}
\begin{align}
&P^{\rm(S)}(\bfk)=\int d^3\bfx\,e^{i\bfk\cdot\bfx}
\bigl\langle e^{-ik\mu\,f\Delta u_z}
\nonumber\\
&\qquad\quad\times
\left\{\delta(\bfr)+f\nabla_zu_z(\bfr)\right\}
\left\{\delta(\bfr')+f\nabla_zu_z(\bfr')\right\}\bigr\rangle,
\label{eq:Pkred_exact}
\end{align}
where $\bfx=\bfr-\bfr'$ and $\langle\cdots\rangle$ is the 
ensemble average. We defined $u_z(\bfr)=-v_z(\bfr)/(aHf)$ and 
$\Delta u_z=u_z(\bfr)-u_z(\bfr')$. The function $f$ is the 
logarithmic derivative of linear growth function $D(z)$ given by 
$f=d\ln D(z)/d\ln a$. This is the exact expression for 
power spectrum in redshift space, and no dynamical information 
for velocity and density fields, i.e., Euler equation and/or continuity 
equation, is invoked in deriving this equation.

In the expression (\ref{eq:Pkred_exact}), 
the power spectrum is written as function of $k$ and $\mu$, and 
is related to the statistical average of real-space quantities in a 
complicated manner, but qualitative effects on clustering amplitude of 
power spectrum are rather clear, i.e., enhancement and damping, well known as 
Kaiser and Finger-of-God effects. The Kaiser effect basically 
comes from the braces in the right hand side of the expression
(\ref{eq:Pkred_exact}), which represents the coherent distortion 
by the peculiar velocity along the line-of-sight direction. In linear 
theory, the relation $u=\delta$ holds and the strength of clustering 
anisotropies is controlled by the growth rate parameter $f$. This is 
the basic reason why the redshift distortion attracts much attention 
as a powerful indicator for growth of structure.  On the other hand, 
Finger-of-God effect roughly comes from 
the factor $e^{-ik\mu\,f\Delta u_z}$ in Eq.~(\ref{eq:Pkred_exact}). 
Due to the randomness of peculiar velocities, de-phasing arises and it 
leads to the suppression of clustering amplitude. The apparent 
reduction of amplitude becomes especially significant 
around the halo forming regions.

Of course, these two effects cannot be separately treated 
in principle,  and a mixture of Kaiser and Finger-of-God effects 
is expected to be significant on trans-linear regime, where 
a tight correlation between velocity and density fields still remains. 
This is of particular importance for the accurate modeling of BAOs. 
Before addressing detailed modeling, however, 
we will first consider currently existing models of redshift distortion, 
and examine how these models fail to 
reproduce the major trends of BAO features in redshift space.

\section{Existing models of redshift distortion}
\label{sec:existing_model}

\subsection{Perturbation theory description}
\label{subsec:PT_model}

Let us first examine the perturbation theory (PT) based model of 
redshift distortion. 
We here specifically deal with the two representative models:
one-loop PT calculations for redshift-space power spectrum from 
standard PT and Lagrangian PT.

The standard PT usually implies a straightforward expansion of the 
the cosmic fluid equations around their linear solution, assuming 
that the amplitudes of density and velocity fields are small. 
This treatment is also applied to the evaluation of redshift-space power 
spectrum (\ref{eq:Pkred_exact}), and the resultant expressions 
for one-loop power spectrum is schematically summarized as 
(see \cite{Heavens:1998es,Matsubara:2007wj} for complete expressions)
\begin{align}
P_{\rm SPT}^{\rm(S)}(k,\mu)=
(1+f\,\mu^2)^2\,P_{\rm lin}(k)+P_{\rm 1\mbox{-}loop}^{\rm(S)}(k,\mu), 
\label{eq:Pk_SPT}
\end{align}
The first term in the right-hand side is the 
linear-order result of the redshift-space power spectrum, and the 
factor $(1+f\,\mu^2)^2$ multiplied by the linear power 
spectrum $P_{\rm lin}$ indicates the enhancement due to the 
Kaiser effect. The second term $P_{\rm 1\mbox{-}loop}^{\rm(S)}$  represents
a collection of the leading-order mode-coupling terms 
called one-loop correction, arising both from the gravitational 
clustering and the redshift distortion. This term is basically of the 
forth order in linear-order density or velocity fields, and 
is roughly proportional to $P_{\rm lin}\Delta^2$ with 
$\Delta^2=k^3P_{\rm lin}/(2\pi^2)$.

On the other hand, the Lagrangian PT description of the 
redshift-space power spectrum is obtained in somewhat different way. 
Intuitively, we rewrite the exact expression (\ref{eq:Pkred_exact}) 
in terms of the  displacement vector, and the perturbative expansion is 
applied to the displacement vector. Although a naive perturbative 
treatment merely reproduces the standard PT result (\ref{eq:Pk_SPT}), 
Ref.~\cite{Matsubara:2007wj} has applied a partial expansion, and some
of the terms has been kept in some non-perturbative ways. The 
resultant expressions for power spectrum in redshift space becomes
\begin{align}
& P_{\rm LPT}^{\rm(S)}(k,\mu)=e^{-k^2\{1+f(f+2)\mu^2\}\sigma_{\rm v,lin}^2}
\nonumber\\
&\,\,\times\left[P_{\rm SPT}^{(S)}(k,\mu)+  
(1+f\,\mu^2)^2\left\{1+f(f+2)\mu^2\right\}k^2\sigma_{\rm v,lin}^2\right], 
\label{eq:Pk_LPT}
\end{align}
where the quantity $\sigma_{\rm v,lin}^2$  is the linear-order estimate of
the one-dimensional velocity dispersion given by
\begin{align} 
\sigma_{\rm v,lin}^2=\frac{1}{3}
\int \frac{d^3\bfq}{(2\pi)^3} \frac{P_{\rm lin}(q,z)}{q^2}.
\label{eq:sigmav_lin}
\end{align} 
The exponential prefactor in Eq.~(\ref{eq:Pk_LPT}) can be regarded 
as the result of non-perturbative treatment, and in redshift space, 
this term accounts for the non-linear damping of the BAOs arising both 
from the gravitational clustering and Finger-of-God effect of redshift 
distortion.

\begin{figure*}
\begin{center}
\includegraphics[width=8.8cm,angle=0]{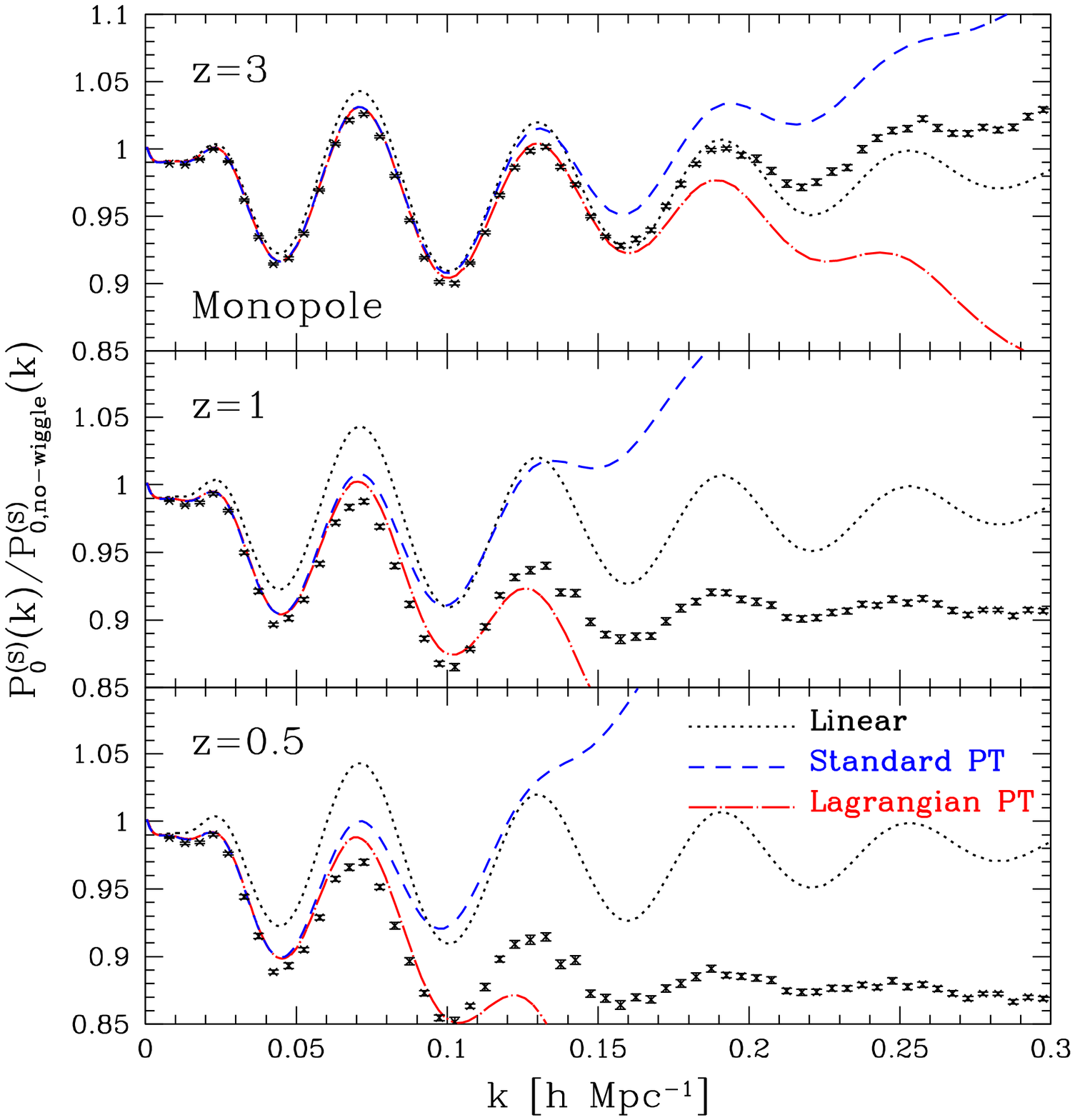}
\includegraphics[width=8.8cm,angle=0]{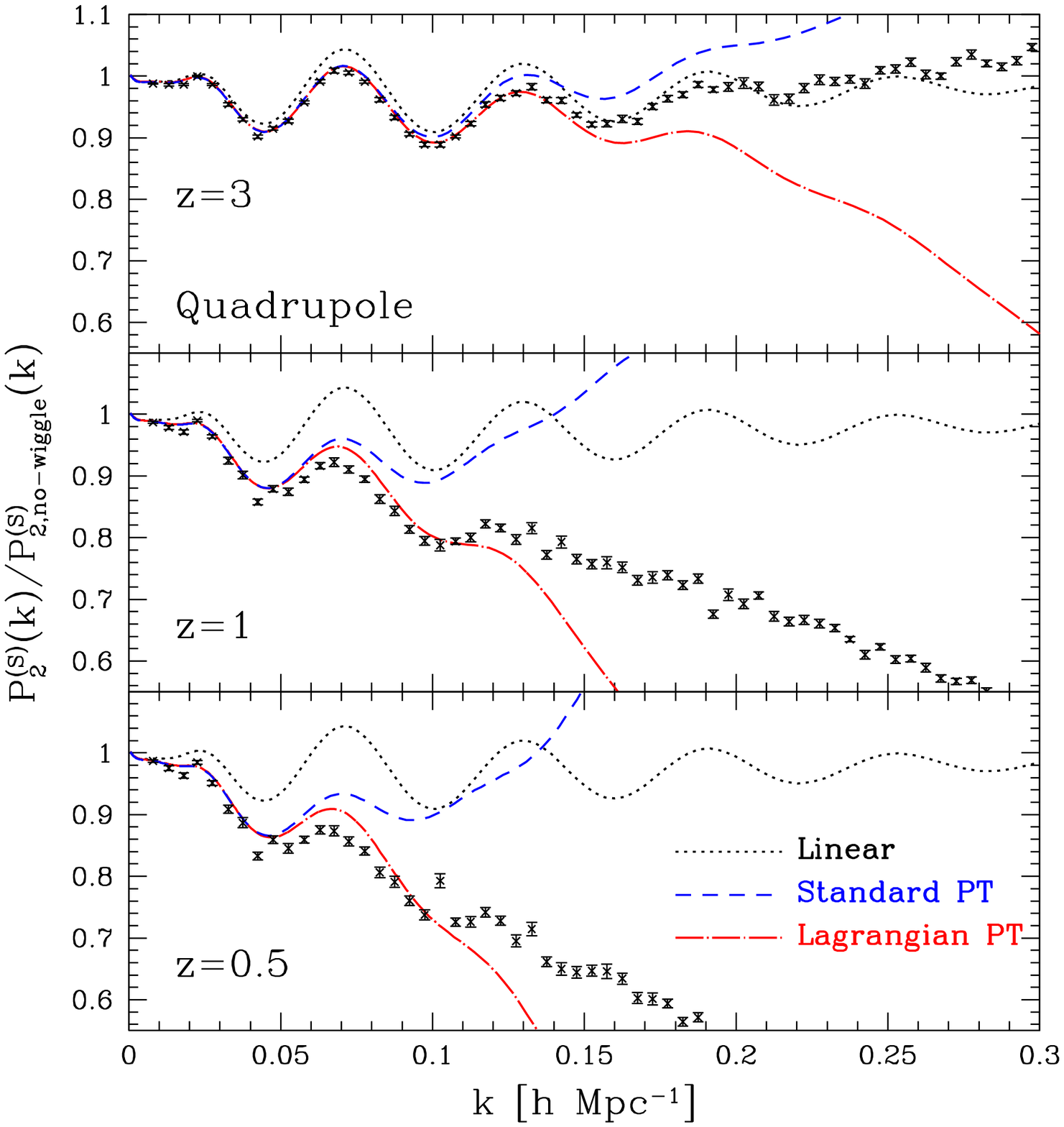}
\end{center}

\vspace*{-0.5cm}

\caption{Ratio of power spectra to smoothed reference spectra in 
  redshift space, 
  $P_\ell^{\rm (S)}(k)/P_{\ell,{\rm no\mbox{-}wiggle}}^{\rm (S)}(k)$. N-body 
  results are taken from the {\tt wmap5} simulations of 
  Ref.~\cite{Taruya:2009ir}. The reference spectrum 
  $P_{\ell,{\rm no\mbox{-}wiggle}}^{\rm (S)}$ is calculated 
  from the no-wiggle approximation of the linear transfer function, and 
  the linear theory of the Kaiser effect is taken into account. 
  Short dashed and dot-dashed lines respectively indicate the results of 
  one-loop PT and Lagrangian PT calculations for redshift-space power spectrum 
  (Eqs.~(\ref{eq:Pk_SPT}) and (\ref{eq:Pk_LPT})). 
\label{fig:ratio_pk_red_PT}}
\end{figure*}

Fig.~\ref{fig:ratio_pk_red_PT} compares the PT based models of 
redshift distortion with N-body simulations of Ref~\cite{Taruya:2009ir}. 
Left and right panels respectively show the 
monopole $(\ell=0)$ and quadrupole $(\ell=2)$ moments of 
power spectrum divided by the smooth reference spectrum at different 
redshifts, $z=3$, $1$ and $0.5$ (from top to bottom). 
The reference spectrum $P_{\ell,{\rm no\mbox{-}wiggle}}^{\rm(S)}(k)$ is 
calculated from the no-wiggle approximation of the linear transfer 
function in Ref.~\cite{Eisenstein:1997ik}, 
taking account of the linear-order result of Kaiser effect. 
The multipole moment of two-dimensional power spectrum is defined by
\begin{equation}
P^{\rm(S)}_\ell(k)=\frac{2\ell+1}{2}\int_{-1}^1d\mu\,P^{\rm(S)}(k,\mu)\,
\mathcal{P}_\ell(\mu),
\end{equation}
with $\mathcal{P}_\ell(\mu)$ being the Legendre polynomials.

As it has been repeatedly stated in the literature 
\cite{Jeong:2006xd,Nishimichi:2008ry,Carlson:2009it,Taruya:2009ir}, 
the standard PT treatment is not sufficiently accurate to describe the 
BAOs. Fig.~\ref{fig:ratio_pk_red_PT} confirms that 
this is indeed true not only in real space, but also in redshift space. 
While the power spectrum amplitude of N-body simulations tends to be 
smaller than that of the linear theory prediction (dotted), 
the predicted amplitude of standard PT generally overestimates the 
N-body results, and it exceeds the linear prediction on small scales. 
Compared to the results in real space,  
the discrepancy between prediction and simulation seems a bit large.  
Contrastingly, in the Lagrangian PT calculation,  
the amplitude of power spectrum is rather suppressed, and a better agreement 
between prediction and simulation is achieved at low-$k$. This is 
due to the exponential prefactor in Eq.~(\ref{eq:Pk_LPT}). 
As a trade-off, however, the predicted amplitude at higher $k$ modes 
largely underestimates the result of N-body simulations. Further, 
a closer look at first peak of BAOs around $k\sim0.05-0.1\,h$Mpc$^{-1}$ 
reveals a small discrepancy, which becomes significant 
as decreasing the redshift and can produce few \% errors in power 
spectrum amplitude.

These results indicate that the existing PT based approaches fail to 
describe the two competitive effects of redshift distortion in the power 
spectrum\footnote{Nevertheless, it should be noted that the 
Lagrangian PT would be still powerful in predicting the two-point 
correlation function around the baryon acoustic peak. In both real and 
redshift spaces, the prediction reasonably recovers the smeared 
peak and trough structures, and it gives a 
better agreement with N-body simulation.}. 
A proper account of these is thus essential in accurately modeling 
BAOs.  

\begin{figure*}
\begin{center}
\includegraphics[width=8.8cm,angle=0]{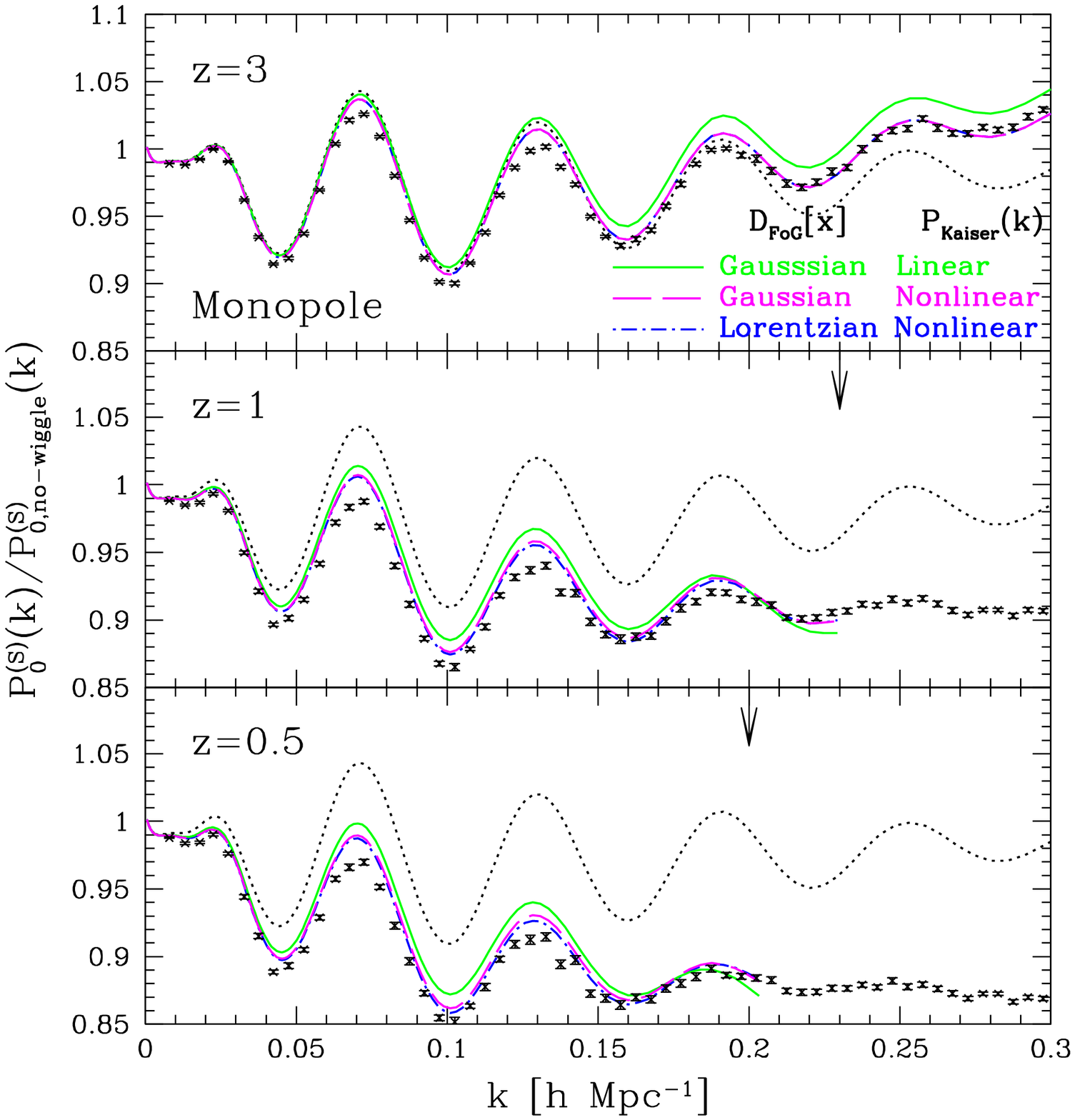}
\includegraphics[width=8.8cm,angle=0]{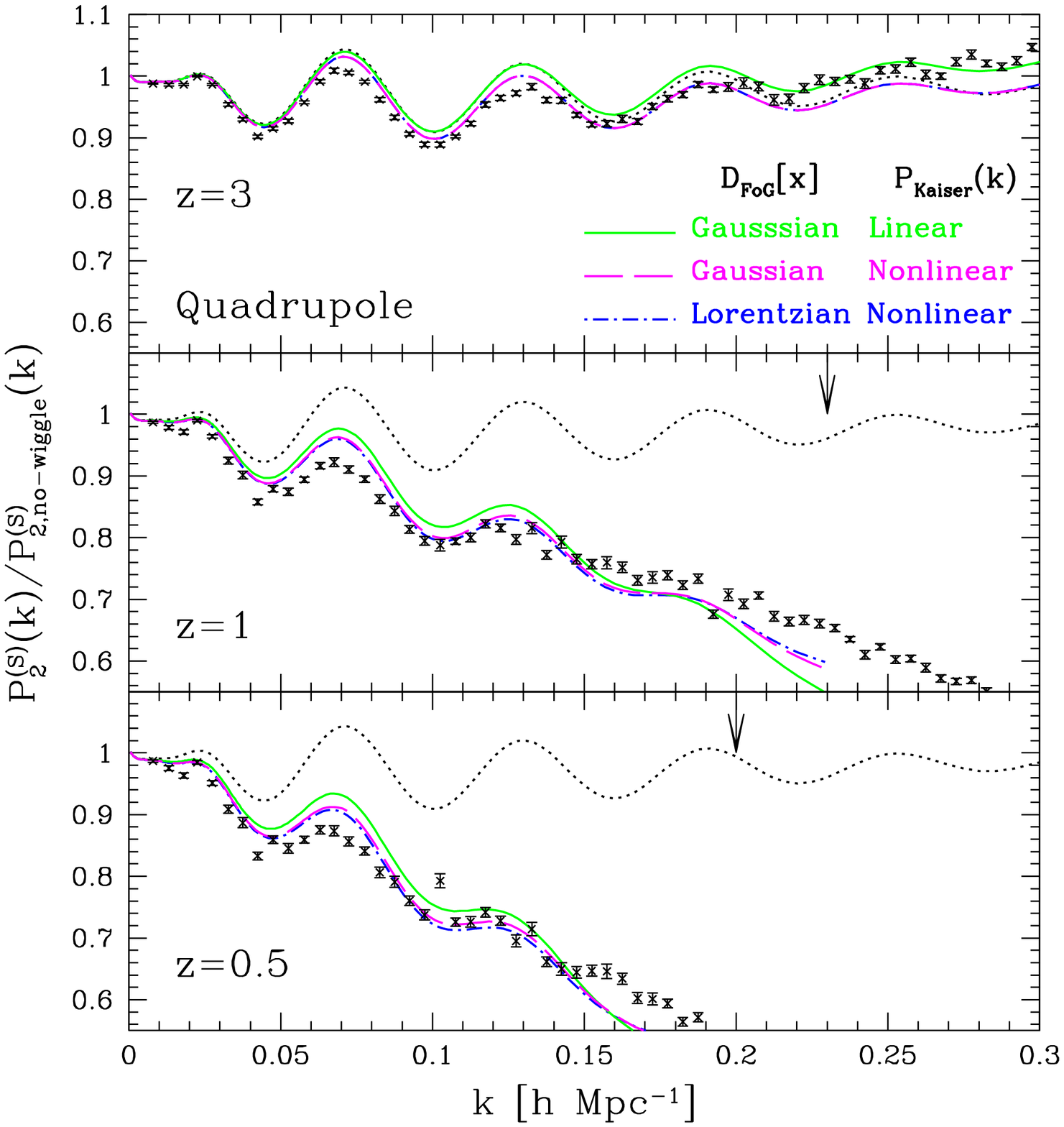}
\end{center}

\vspace*{-0.5cm}

\caption{Same as in Fig.~\ref{fig:ratio_pk_red_PT}, but we here plot 
  the results of phenomenological model predictions. The three
  different predictions depicted as solid, dashed, dot-dashed lines are
  based on the phenomenological model of redshift distortion 
  (\ref{eq:model_Psk}) with various choices of Kaiser and Finger-of-God 
  terms (Eqs.(\ref{eq:model_Kaiser}) and (\ref{eq:model_FoG})). 
  Left panel shows the monopole power spectra ($\ell=0$), and 
  the right panel shows the quadrupole spectra ($\ell=2$).  
  In all cases, one-dimensional velocity dispersion 
  $\sigma_{\rm v}$ was determined 
  by fitting the predictions to the N-body simulations.  In each 
  panel, vertical arrow indicates the maximum wavenumber $k_{1\%}$ 
  for improved PT prediction including up to the second-order Born 
  approximation (see Eq.~(\ref{eq:k_limit}) for definition). 
\label{fig:ratio_pk_red_phenom}}
\end{figure*}

\subsection{Phenomenological model description}
\label{subsec:phenom_model}

Next consider the phenomenological models of redshift distortion, 
which have been originally introduced to explain 
the observed power spectrum on small scales. Although the 
relation between the model and exact expression (\ref{eq:Pkred_exact}) 
is less clear,  for most of the models frequently used in the 
literature, the redshift-space power spectrum is expressed in the form as 
(e.g., \cite{Scoccimarro:2004tg,Percival:2008sh,Cole:1994wf,
Peacock:1993xg,Park:1994fa,Ballinger:1996cd,Magira:1999bn})
\begin{equation}
P^{\rm(S)}(k,\mu) = \DFoG[k\mu f\,\sigmav] \,\PKaiser(k,\mu), 
\label{eq:model_Psk}
\end{equation}
where the term $\PKaiser(k,\mu)$ represents the Kaiser effect, and the 
term $\DFoG[k\,\mu\,f\,\sigmav]$ indicates the damping function which 
mimics the Finger-of-God effect. The quantity $\sigmav$ is the 
one-dimensional velocity dispersion defined by 
$\sigmav^2=\langle u_z^2(0)\rangle$. 
The variety of the functional form for $\PKaiser(k,\mu)$ 
and $\DFoG[k\,\mu\,f\,\sigmav]$ are summarized as follows. 
\begin{figure}
\begin{center}
\includegraphics[width=8.5cm,angle=0]{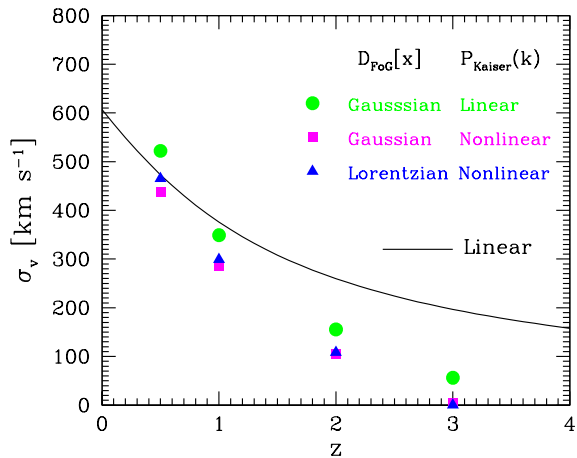}
\end{center}

\vspace*{-1.8cm}

\caption{Redshift evolution of velocity dispersion $\sigma_{\rm v}$ 
  determined by fitting the predictions of monopole and quadrupole power 
  spectra to the N-body results. 
  While the solid lines represent the linear theory prediction, 
  the symbols indicate the results obtained by fitting 
  models of redshift distortion with various choices of Kaiser and 
  damping terms (see Fig.~\ref{fig:ratio_pk_red_phenom}). 
\label{fig:sigma_v}}
\end{figure}

The Kaiser effect has been first recognized from the linear-order 
calculations \cite{Kaiser:1987qv}, 
from which the enhancement factor $(1+f\,\mu^2)^2$ is 
obtained (see Eq.~(\ref{eq:Pk_SPT})). As a simple description for the Kaiser 
effect, one may naively multiply the non-linear matter power spectrum 
by this factor, just by hand. 
Recently, proper account of the non-linear effect has 
been discussed \cite{Scoccimarro:2004tg,Percival:2008sh}, 
and non-linear model of Kaiser effect has been 
proposed using the real-space power spectra. Thus, we have 
\begin{align}
&\PKaiser(k,\mu)
\nonumber\\
&\,\,=
\left\{
\begin{array}{lcl}
(1+f\mu^2)^2\Pdd(k)&;&\mbox{linear}
\\
\\
\Pdd(k)+2f\,\mu^2\,\Pdv(k)+
f^2\,\mu^4\,\Pvv(k)&;&\mbox{non-linear}
\end{array}
\right.
\nonumber\\
\label{eq:model_Kaiser}
\end{align}
Here, the spectra $\Pdd$, $\Pvv$, and $\Pdv$ denote the 
auto power spectra of density and velocity divergence, and their cross power 
spectrum, respectively. The velocity divergence $\theta$ is defined by 
$\theta\equiv\nabla \bfu=-\nabla \bfv/(aHf)$ \footnote{The sign convention of 
the definition of velocity divergence $\theta$ differs from that of Refs.~
\cite{Taruya:2007xy,Taruya:2009ir}, but is equivalent to the one in 
Refs.~\cite{Scoccimarro:2004tg,Crocce:2005xy,Crocce:2005xz,Crocce:2007dt}.}.

On the other hand, the functional form of the damping term can 
be basically modeled from the distribution function of one-dimensional 
velocity. Historically, it is characterized by Gaussian or exponential 
function (e.g., \cite{Peacock:1993xg,Park:1994fa,Ballinger:1996cd,
Magira:1999bn}), which lead to 
\begin{eqnarray}
\DFoG[x]=\left\{
\begin{array}{lcl}
\exp(-x^2)&;& \mbox{Gaussian}
\\
\\
1/(1+x^2)&;& \mbox{Lorentzian}
\end{array}
\right.
\label{eq:model_FoG}
\end{eqnarray}
Note that there is analogous expression for exponential distribution, 
i.e., $\DFoG[x]=1/(1+x^2/2)^2$ \cite{Cole:1994wf},  
but the resultant power spectrum is quite similar to the one
adopting the Lorentzian form 
for the range of our interest, $x\lesssim1$. 
Since the Finger-of-God effect is thought to be a 
fully non-linear effect, which mostly comes from the virialized random 
motion of the mass (or galaxy) residing at a halo,  
the prediction of $\sigmav$ seems rather difficult.  
Our primary purpose is to model the shape and structure of 
acoustic feature in the power spectrum, and the precise form of the 
damping is basically irrelevant. We thus regard $\sigmav$ as a 
free parameter, and determine it by fitting the predictions to the 
simulations or observations.

Fig.~\ref{fig:ratio_pk_red_phenom} compares the phenomenological 
models of redshift distortion with combination of 
Eqs.~(\ref{eq:model_Kaiser}) and (\ref{eq:model_FoG}) with 
N-body simulations. In computing the redshift-space power spectrum from the 
phenomenological models,  
we adopt the improved PT treatment by Refs.~\cite{Taruya:2007xy,Taruya:2009ir}, 
and the analytic results including the corrections up to the 
second-order Born approximation are used to obtain 
the three different power spectra 
$\Pdd$, $\Pdv$ and $\Pvv$. The accuracy of the improved PT treatment 
has been checked in details by Refs.~\cite{Taruya:2009ir}, and it 
has been shown that 
the predictions of $\Pdd$ reproduce the N-body results quite well 
within $1\%$ accuracy below the wavenumber $k_{1\%}$, 
indicated by the vertical arrows 
in Fig.~\ref{fig:ratio_pk_red_phenom}. This has been 
calibrated from a proper comparison between N-body and PT results
and is empirically characterized by solving the following equation 
\cite{Nishimichi:2008ry,Taruya:2009ir}:
\begin{align}
\frac{k_{1\%}^2}{6\pi^2}\int_0^{k_{1\%}} dq\,P_{\rm lin}(q;z)=C
\label{eq:k_limit}
\end{align}
with $C=0.7$ and $P_{\rm lin}$ being linear matter spectrum. 
Note that the $1\%$ accuracy of the improved PT prediction 
at $z=3$ has reached at $k\sim0.47h$Mpc$^{-1}$, outside the plot range. 
We basically use this criterion to determine $\sigmav$, 
and fit the predictions of both monopole and quadrupole spectra 
to the N-body results in the range $0\leq k\leq k_{1\%}$.

Since we allow $\sigmav$ to vary as a free parameter, 
the overall behaviors of the model predictions reproduce with N-body 
results, and the differences between model predictions are basically small  
compared to the results in the PT description. 
However, there still exist small but non-negligible 
discrepancies between N-body results 
and model predictions, which are statistically significant, and are 
comparable or exceed the expected errors in upcoming BAO measurements 
\cite{Taruya:2009ir}. 
Although the agreement is somehow improved when 
we adopt the non-linear model of $\PKaiser$,  
there still remains discrepancies of few \% in monopole 
and 5 \% in quadrupole moments of power spectrum amplitudes.  
These are irrespective of the choice of damping function $\DFoG$.

Furthermore, the fitted results of $\sigmav$ show somewhat peculiar behavior. 
Fig.~\ref{fig:sigma_v} plots the fitted values of $\sigmav$ as function 
of redshift (symbols), which significantly deviate from linear theory 
prediction (solid line) as increasing the redshifts. This is 
in contrast with a naive expectation, and indicates 
that the model based on the expression (\ref{eq:model_Psk}) 
misses something important, and needs to be reconsidered.

\section{Improved model prediction}
\label{sec:improved_model}

\subsection{Derivation}
\label{subsec:derivation}

Comparison in previous section reveals that even in the 
models with fitting parameter, 
a small but non-negligible discrepancy appears at the scales of BAOs,  
where the choice of the damping function $\DFoG[x]$ does not sensitively 
affect the predictions. This implies that there exists missing terms arising 
from the non-linear mode coupling between 
density and velocity fields, and those corrections alter 
the acoustic feature in redshift-space power spectrum. 
In this section, starting with the exact expression (\ref{eq:Pkred_exact}), 
we derive non-linear corrections, which are relevant to explain 
the modulation of acoustic features in redshift space.

First recall that the expression (\ref{eq:Pkred_exact}) is written in the 
form as 
\begin{equation}
P^{\rm(S)}(k,\mu)=\int d^3\bfx\,e^{i\,\bfk\cdot\bfx}
\bigl\langle e^{j_1A_1}A_2A_3\bigr\rangle, 
\label{eq:Pks_exact}
\end{equation}
where the quantities $j_1$, $A_i\,(i=1,2,3)$ are respectively given by 
\begin{align}
&j_1= -i\,k\mu f,\nonumber\\
&A_1=u_z(\bfr)-u_z(\bfr'),\nonumber\\
&A_2=\delta(\bfr)+f\,\nabla_zu_z(\bfr),\nonumber\\
&A_3=\delta(\bfr')+f\,\nabla_zu_z(\bfr').\nonumber
\end{align}
We shall rewrite the ensemble average 
$\langle e^{j_1A_1}A_2A_3\rangle$ in terms of the cumulants. 
To do this, we use the relation between the cumulant and 
moment generating functions. For the stochastic vector field 
$\bfA=\{A_1,A_2,A_3\}$, we have 
(e.g., \cite{Scoccimarro:2004tg, Matsubara:2007wj}): 
\begin{equation}
\langle e^{\bfj\cdot\bfA}\rangle=
\exp \left\{\langle e^{\bfj\cdot\bfA}\rangle_c\right\}
\end{equation}
with $\bfj$ being arbitrary constant vector, $\bfj=\{j_1,j_2,j_3\}$. 
Taking the derivative twice with 
respect to $j_2$ and $j_3$, and we then set $j_2=j_3=0$. We obtain 
\cite{Scoccimarro:2004tg}
\begin{align}
&\langle e^{j_1A_1}A_2A_3\rangle=
\exp \left\{\langle e^{j_1A_1}\rangle_c\right\}
\nonumber\\
&\qquad\times
\left[\langle e^{j_1A_1}A_2A_3 \rangle_c+ 
\langle e^{j_1A_1}A_2\rangle_c \langle e^{j_1A_1}A_3 \rangle_c \right].
\end{align}
Substituting this into Eq.(\ref{eq:Pks_exact}), we arrive at
\begin{align}
&P^{\rm(S)}(k,\mu)=\int d^3\bfx \,\,e^{i\bfk\cdot\bfx}\,\,
\exp \left\{\langle e^{j_1A_1}\rangle_c\right\}
\nonumber\\
&\quad\quad
\times\left[\langle e^{j_1A_1}A_2A_3 \rangle_c+ 
\langle e^{j_1A_1}A_2\rangle_c \langle e^{j_1A_1}A_3 \rangle_c \right].
\label{eq:Pkred_exact2}
\end{align}
This expression clearly reveals the coupling between density and velocity 
fields associated with Kaiser and Finger-of-God effects. 
In addition to the prefactor $\exp \left\{\langle e^{j_1A_1}\rangle_c\right\}$, 
the ensemble averages over the quantities $A_2$ and $A_3$ 
responsible for the Kaiser effect 
all includes the exponential factor $e^{j_1A_1}$, which can produce 
a non-negligible correlation between density and velocity.

Comparing Eq.~(\ref{eq:Pkred_exact2}) with the expression 
(\ref{eq:model_Psk}) with (\ref{eq:model_Kaiser}) and (\ref{eq:model_FoG}),  
we deduce that the phenomenological models discussed in 
Sec.~\ref{subsec:phenom_model} miss something important, 
and are derived based on several assumptions or treatments:  

\begin{itemize}
\item In the integrand of Eq.~(\ref{eq:Pkred_exact2}), while taking 
  the limit $j_1\to0$ in the bracket, we keep $j_1\neq0$ in the exponent 
  of the prefactor. 
\item For cumulants 
  $\langle A_1^n\rangle_c=\langle[u_z(\bfr)-u_z(\bfr')]^n\rangle_c$ 
  of any integer value $n$, 
the spatial correlations between different positions are ignored, 
and the non-vanishing cumulants are assumed to be expressed as 
$\langle A_1^n\rangle_c\simeq2\langle u_z^n\rangle_c= 2c_n\,\sigmav^n$ 
for even number of $n$, with $c_n$ being constants. 
\item To further obtain the Gaussian or Lorentzian forms of the 
damping function $\DFoG[x]$, we assume that 
the conditions, $c_n=0$ except for $c_2=1$, or, 
$c_{2n}=(2n-1)!$ and $c_{2n-1}=0$, are fulfilled. 
\end{itemize}

In the above, the last two conditions play a role for specifying the 
damping function, and they mainly affect the broadband shape of the power 
spectrum. On the other hand, the first condition leads to the expression 
of $\PKaiser(k)$, which can add the most dominant contribution to the 
acoustic feature in power spectrum. Since the choice of the damping function 
(\ref{eq:model_FoG}) is presumably a minor source for discrepancies between 
the model predictions and simulations, taking the limit $j_1\to0$ in the 
bracket would be the main reason for discrepancy.  In this respect, 
the terms involving the exponential factor can produce additional 
contributions to the spectrum $\PKaiser(k)$, which are 
responsible for explaining the modulated acoustic peak and trough 
structure in redshift space.

Let us now derive the corrections to $\PKaiser(k)$. To do this,  
we keep the last two conditions, and perturbatively 
treat the terms inside the bracket of Eq.~(\ref{eq:Pkred_exact2}).  
This treatment is reasonable, because the modification of acoustic 
features should be small for relevant scales of BAOs. 
On the other hand, the factor 
$\exp\{\langle e^{j_1A_1}\rangle_c\}$ is most likely affected 
by the virialized random motion of the mass around halos, 
and seems difficult to treat it perturbatively. 
Here, regarding the quantity $j_1$ as a small expansion parameter, we 
perturbatively expand the terms in the bracket of the integrand. Up to 
the second order in $j_1$, we have
\begin{align}
&\langle e^{j_1A_1}A_2A_3 \rangle_c+
\langle e^{j_1A_1}A_2\rangle_c \langle e^{j_1A_1}A_3 \rangle_c 
\nonumber\\
&\,\,\,\simeq\langle A_2A_3\rangle + j_1\langle A_1A_2A_3\rangle_c
\nonumber\\
&\quad
+j_1^2\Bigl\{ \frac{1}{2}\,\langle A_1^2A_2A_3\rangle_c 
+\langle A_1A_2\rangle_c\langle A_1A_3\rangle_c
\Bigr\}+\mathcal{O}(j_1^3).
\end{align}
In the above, the term $\langle A_1^2A_2A_3\rangle_c$ turns out to be
higher order when we explicitly compute it employing the perturbation 
theory calculation, and is roughly proportional to 
$\mathcal{O}(P_{\rm lin}^3)$. We thus drop the 
higher-order contribution, and collect the leading and next-to-leading 
order contributions. Then, Eq.~(\ref{eq:Pkred_exact2}) can be recast as
\begin{align}
&P^{\rm(S)}(k,\mu)=\DFoG[k\mu\,f\,\sigmav]\,\Bigl\{\Pdd(k)
+2\,f\,\mu^2\,\Pdv(k)
\nonumber\\
&\qquad+f^2\,\mu^4\,\Pvv(k)+A(k,\mu)+B(k,\mu)
\Bigr\}. 
\label{eq:new_model}
\end{align}
Here, we replaced the exponential prefactor 
$\exp\{\langle e^{j_1A_1}\rangle_c\}$ with the damping function $\DFoG$.   
The corrections $A$ and $B$ are respectively given by 
\begin{eqnarray}
  \label{eq:correction}
  A(k,\mu)&=& j_1\,\int d^3\bfx \,\,e^{i\bfk\cdot\bfx}\,\,
\langle A_1A_2A_3\rangle_c,
\nonumber\\
  B(k,\mu)&=& j_1^2\,\int d^3\bfx \,\,e^{i\bfk\cdot\bfx}\,\,
\langle A_1A_2\rangle_c\,\langle A_1A_3\rangle_c.
\nonumber
\end{eqnarray}
In terms of the basic quantities of density $\delta$ and velocity divergence 
$\theta=-\nabla\bfv/(aHf)$, they are rewritten as 
\begin{align}
&A(k,\mu)= (k\mu\,f)\,\int \frac{d^3\bfp}{(2\pi)^3} \,\,\frac{p_z}{p^2}
\nonumber\\
&\qquad\quad\times
\left\{B_\sigma(\bfp,\bfk-\bfp,-\bfk)-B_\sigma(\bfp,\bfk,-\bfk-\bfp)\right\},
\label{eq:A_term}
\\
&B(k,\mu)= (k\mu\,f)^2\int \frac{d^3\bfp}{(2\pi)^3} F(\bfp)F(\bfk-\bfp)\,\,;
\label{eq:B_term}
\\
&\qquad\quad F(\bfp)=\frac{p_z}{p^2}
\left\{ \Pdv(p)+f\,\frac{p_z^2}{p^2}\,\Pvv(p)\,\right\},
\nonumber
\end{align}
where the function $B_\sigma$ is the cross bispectra defined by 
\begin{align}
&\left\langle \theta(\bfk_1)
\left\{\delta(\bfk_2)+f\,\frac{k_{2z}^2}{k_2^2}\theta(\bfk_2)\right\}
\left\{\delta(\bfk_3)+f\,\frac{k_{3z}^2}{k_3^2}\theta(\bfk_3)\right\}
\right\rangle
\nonumber\\
&\quad\qquad
=(2\pi)^3\delta_D(\bfk_1+\bfk_2+\bfk_3)\,B_\sigma(\bfk_1,\bfk_2,\bfk_3).
\label{eq:def_B_sigma}
\end{align}

In deriving the expression (\ref{eq:new_model}),  while we employed 
the low-$k$ expansion, 
we do not assume that the terms $A_i$ themselves are entirely small. 
In this sense, the expressions (\ref{eq:new_model}), (\ref{eq:A_term}) and
(\ref{eq:B_term}) still have some non-perturbative properties, 
although the new corrections $A$ and $B$ 
neglected in the previous phenomenological models are expected to be small, and 
can be treated perturbatively. In Appendix 
\ref{appendix:PT_calc_correction}, based on the standard PT treatment, 
we summarize the perturbative expressions for the corrections 
(\ref{eq:A_term}) and (\ref{eq:B_term}), in which the 
three-dimensional integrals are reduced to the sum of the 
one- and two-dimensional integrals.

\begin{figure}[t]
\begin{center}
\vspace*{-1.5cm}

\includegraphics[width=8.8cm,angle=0]{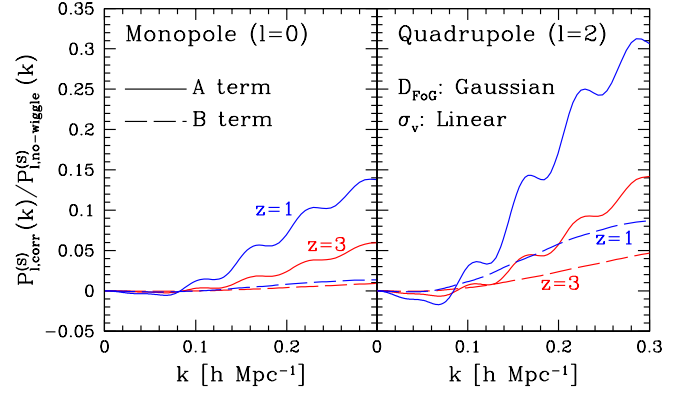}
\end{center}

\vspace*{-1.5cm}

\caption{Contributions of power spectrum corrections coming from 
  the $A$ and $B$ terms divided by the smooth reference power spectrum, 
  $P^{\rm(S)}_{\ell,{\rm corr}}(k)/P^{\rm(S)}_{\ell,{\rm no\mbox{-}wiggle}}(k)$ 
  (Eq.~(\ref{eq:pkred_corr})).  
  We adopt the Gaussian form of the damping function $\DFoG$ with 
  $\sigmav$ computed from linear theory (see Eq.(\ref{eq:sigmav_lin})). 
  Left and right panels respectively show the monopole and 
  quadrupole power spectra at redshifts $z=3$ and $1$. 
\label{fig:ratio_pk_corr}}
\end{figure}
To see the significance of the newly 
derived terms $A$ and $B$, we evaluate the monopole and quadrupole 
contributions to the functions defined by
\begin{align}
P_{\ell,{\rm corr}}^{\rm(S)}(k) \equiv
\frac{2\ell+1}{2} \int_{-1}^1 d\mu\,\,
\DFoG(k\mu f\sigmav)\, \left\{
\begin{array}{c}
A(k,\mu) 
\\ \\
B(k,\mu)
\end{array}
\right\}. 
\label{eq:pkred_corr}
\end{align}
The results are then plotted in 
Fig.~\ref{fig:ratio_pk_corr}, divided by the smoothed reference spectrum, 
$P_{\ell,{\rm no\mbox{-}wiggle}}^{\rm(S)}(k)$. In plotting the 
results, we specifically assume the Gaussian form of $\DFoG$, and adopt the 
linear theory to estimate $\sigmav$ (see Eq.~(\ref{eq:sigmav_lin})).

The corrections coming from the $A$ term show oscillatory behaviors, and 
tend to have a larger amplitude than those from the $B$ term. 
While the corrections from the $B$ term are basically smooth and small, 
they still yield a non-negligible contribution, especially for 
quadrupole power spectrum. 
Although the actual contributions of these corrections 
to the total power spectrum are determined by the fitting parameter $\sigmav$, 
and thus the resultant amplitudes shown in Fig.~\ref{fig:ratio_pk_corr} 
do not simply reflect the correct amplitudes, 
the new corrections $A$ and $B$ can definitely give an important 
contribution to the acoustic feature in power spectrum.

\begin{figure*}
\begin{center}
 \includegraphics[width=8.8cm,angle=0]{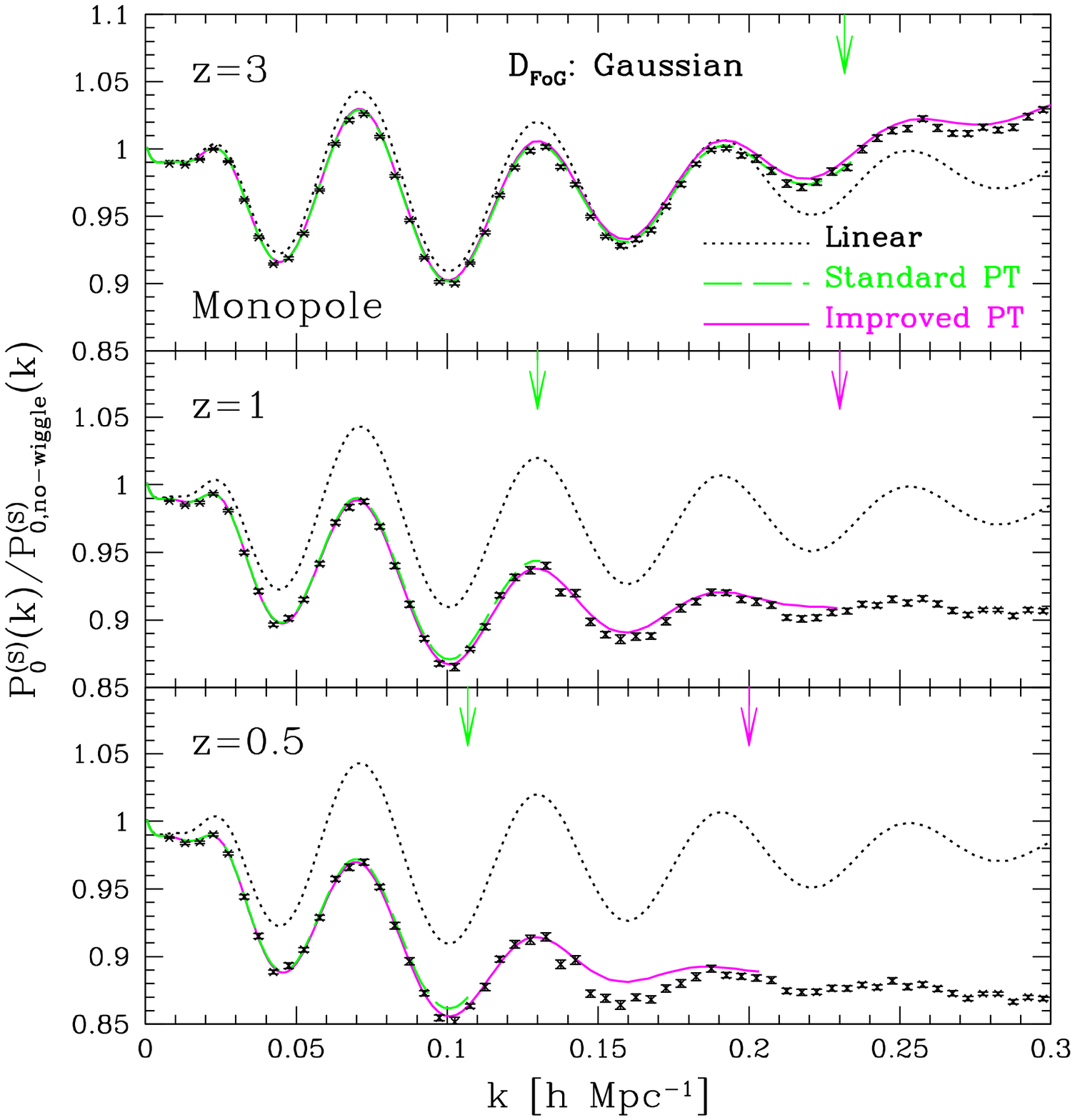}
 \includegraphics[width=8.8cm,angle=0]{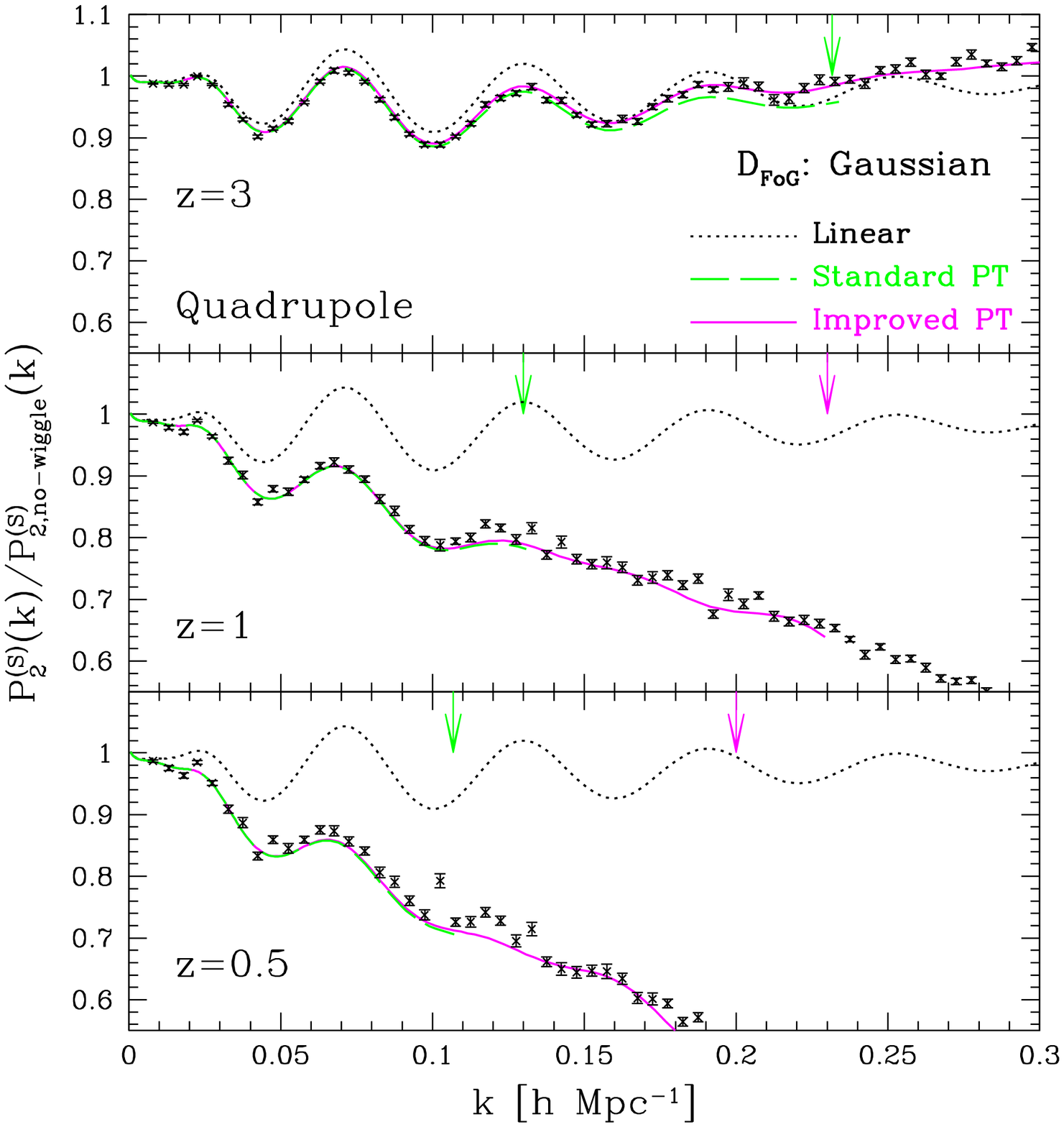}
\end{center}

\vspace*{-0.5cm}

\caption{Same as in Fig.~\ref{fig:ratio_pk_red_phenom}, but we here adopt new 
  model of redshift distortion (\ref{eq:new_model}). 
  Solid and dashed lines represent the predictions for which the spectra 
  $\Pdd$, $\Pdv$ and $\Pvv$ are obtained from the improved PT including 
  the correction up to the second-order Born correction, and one-loop 
  calculations of the standard PT, respectively. 
  In both cases, the corrections $A$ and $B$ given in 
  Eqs. (\ref{eq:A_term}) and (\ref{eq:B_term}) are calculated from 
  standard PT results (see Appendix \ref{appendix:PT_calc_correction}). 
  The vertical arrows indicate the maximum wavenumber $k_{1\%}$ defined 
  in Eq.~(\ref{eq:k_limit}), for standard PT and improved PT 
  (from left to right).
\label{fig:ratio_pk_red2}}
\end{figure*}

Finally, it is interesting to note that while the new formula for 
redshift-space power spectrum (\ref{eq:new_model}) would be applicable 
to the non-linear regime where the standard PT calculation breaks down, 
the resultant expression itself is similar to 
the one for redshift-space power spectrum in the one-loop standard PT. 
The one-loop power spectrum in redshift space, 
$P^{\rm(S)}_{\rm SPT}(k,\mu)$ given at Eq.~(\ref{eq:Pk_SPT}), 
can be formally recast as 
\begin{align}
&P^{\rm(S)}_{\rm SPT}(k,\mu)= 
\left\{1-(k\mu f \sigma_{\rm v, lin})^2\right\}\,\left\{\Pdd(k)+
2f\,\mu^2\Pdv(k)\right.
\nonumber\\
&\quad\left.+f^2\mu^4\Pvv(k)\right\}+A(k,\mu)+B(k,\mu)+C(k,\mu).
\label{eq:1loop_pkred}
\end{align}
Note that each term in the above expression should be 
consistently evaluated using the perturbative solutions up to 
the third order in $\delta$ and $\theta$, and as a result, 
only the leading-order corrections just proportional to $P_{\rm lin}\Delta^2$ 
(or 
equivalently the forth order in $\delta^{(1)}$) are included in 
the one-loop power spectrum. Here, the function $C$ is defined by 
\begin{align}
C(k,\mu)&=(k\mu\,f)^2\int\frac{d^3\bfp d^3\bfq}{(2\pi)^3}\,
\delta_{\rm D}(\bfk-\bfp-\bfq)\,\frac{\mu_p^2}{p^2}\Pvv(p)
\nonumber\\
&\qquad\times
\left\{\Pdd(q)+2\,f\,\mu_q^2\,\Pdv(q)+f^2\,\mu_q^4\,\Pvv(q)\right\}
\nonumber\\
&\simeq  (k\mu\,f)^2\int\frac{d^3\bfp d^3\bfq}{(2\pi)^3}\,
\delta_{\rm D}(\bfk-\bfp-\bfq)\frac{\mu_p^2}{p^2}
\left(1+f\,\mu_q^2\right)^2
\nonumber\\
&\qquad\qquad \times \,P_{\rm lin}(p)P_{\rm lin}(q)
\end{align}
with $\mu_p=p_z/|\bfp|$ and $\mu_q=q_z/|\bfq|$. The second equality 
is valid for the one-loop PT calculation. Hence, 
if we adopt either of Lorentzian or Gaussian form in 
Eq.~(\ref{eq:model_FoG}) and just expand it in powers of its argument, 
the new formula (\ref{eq:new_model}) reduces to 
the one-loop result (\ref{eq:1loop_pkred}) just dropping the term $C$.

The $C$ term is originated from the spatial correlation of the velocity 
field, and is obtained through the low-$k$ expansion of the 
exponential prefactor $\exp\{\langle e^{j_1A_1}\rangle_c\}$ in 
Eq.~(\ref{eq:Pkred_exact2}). For the scales of BAOs, 
the $C$ term monotonically increases the amplitude of power spectrum, 
and it does not alter the acoustic structure drastically. 
Indeed, our several examinations reveal that 
the effect of this can be effectively absorbed into the damping 
function $D[k\mu f\sigmav]$ with varying the velocity dispersion $\sigmav$. 
Rather, the main drawback of the standard PT expression (\ref{eq:1loop_pkred}) 
comes from a naive expansion of all the terms in the exact formula 
(\ref{eq:Pkred_exact}), which fails to describe the delicate balance between 
the Finger-of-God damping and the enhancement from Kaiser effect and 
non-linear gravitational growth. As we will see in next subsection, both 
keeping the damping term $\DFoG$ 
and including the corrections $A$ and $B$ seem essential, and with this 
treatment, even the standard PT calculation of the power spectrum 
can give a excellent result which reproduces the N-body simulations 
fairly well.

\subsection{Comparison with N-body simulations}
\label{subsec:comparison}

We now compare the new prediction of redshift-space  
power spectra with the result of N-body simulations. 
Fig.~\ref{fig:ratio_pk_red2} shows the monopole (left) and quadrupole 
(right) power spectra divided by their smooth reference spectra. 
The analytical predictions based on the model 
(\ref{eq:new_model}) are plotted adopting the 
Gaussian form of the Finger-of-God term $\DFoG[kf\mu\sigmav]$, 
and the velocity dispersion $\sigmav$ is determined by fitting 
the predictions to the N-body results. In computing the predictions,  
the $A$ and $B$ terms are calculated from 
the one-loop standard PT results in Appendix 
\ref{appendix:PT_calc_correction}, while the spectra 
$\Pdd$, $\Pdv$ and $\Pvv$ are obtained 
from improved PT in solid lines, and from standard PT in dashed lines.

\begin{figure}[t]

\vspace*{-2.0cm}

\begin{center}
\includegraphics[width=8.6cm,angle=0]{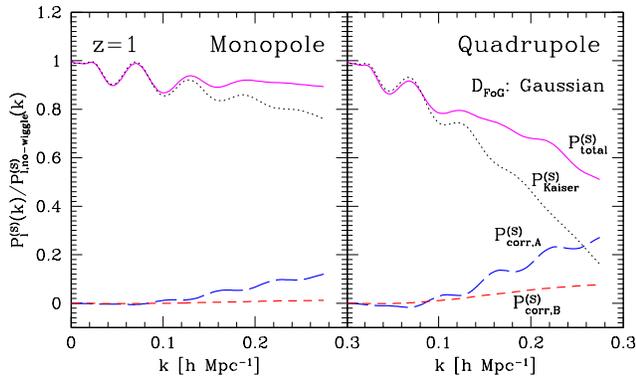}
\end{center}

\vspace*{-1.5cm}

\caption{Contribution of each term in redshift-space power spectrum.  
  For monopole ($\ell=0$, left) and quadrupole ($\ell=2$, right) 
  spectra of the improved model prediction at $z=1$ shown in 
  solid lines of Fig.~\ref{fig:ratio_pk_red2},  we divide the total 
  power spectrum $P_{\rm total}^{\rm(S)}$ (solid) into the three pieces 
  as $P_{\rm total}^{\rm(S)}=P_{\rm Kaiser}^{\rm(S)}+P_{\rm corr,A}^{\rm(S)}
  +P_{\rm corr,B}^{\rm(S)}$, and each contribution is separately plotted 
  dividing by smoothed reference spectra, 
  $P_{\ell,{\rm no\mbox{-}wiggle}}^{\rm(S)}$.   
  Here, the spectrum $P_{\rm Kaiser}^{\rm(S)}$ (dotted) is the contribution of 
  non-linear Kaiser term (\ref{eq:model_Kaiser}) 
  convolved with the Finger-of-God damping $\DFoG$, and the corrections 
  $P_{\rm corr,A}^{\rm(S)}$ and $P_{\rm corr,B}^{\rm(S)}$ are those given 
  by Eq.~(\ref{eq:pkred_corr}). 
\label{fig:pk_contribution}}
\end{figure}

Compared to Figs.~\ref{fig:ratio_pk_red_PT} and 
\ref{fig:ratio_pk_red_phenom}, 
the agreement between N-body simulations and 
predictions depicted as solid lines   
becomes clearly improved, and the prediction including the corrections 
faithfully traces the N-body trends of acoustic feature, especially 
around $k=0.05\sim0.15h$Mpc$^{-1}$, where the phenomenological model
shows a few \% level discrepancy. A remarkable point is that 
a reasonable agreement basically holds over the range 
below the critical wavenumber 
$k_{1\%}$ calibrated in real space (vertical arrows, 
Eq. (\ref{eq:k_limit}) for definition). This is also true for the case 
adopting one-loop standard PT to compute $\Pdd$, $\Pdv$ and $\Pvv$ 
(dashed lines),  and the range of agreement is wider than that  
of the existing PT-based models in Sec.~\ref{subsec:PT_model}.

In Fig.~\ref{fig:pk_contribution}, to see the significance of 
the contributions from corrections $A$ and $B$, 
we divide the improved PT prediction of power spectra $P^{\rm(S)}(k)$ at $z=1$ 
into the three pieces as $P_{\rm Kaiser}^{\rm(S)}$, $P_{\rm corr,A}^{\rm(S)}$ 
and $P_{\rm corr,B}^{\rm(S)}$, which are separately plotted as dotted, 
long-dashed, and short dashed lines, respectively. 
The power spectrum $P_{\rm Kaiser}^{\rm(S)}$ is the contribution of 
the non-linear Kaiser term given in Eq.~(\ref{eq:model_Kaiser}),  
convolved with the damping function $\DFoG$. 
The spectra $P_{\rm corr,A}^{\rm(S)}$ and $P_{\rm corr,B}^{\rm(S)}$ represent 
the actual contributions of the corrections $A$ and $B$ 
defined by Eq.~(\ref{eq:pkred_corr}), with fitted value of $\sigmav$.  
The corrections $A$ and $B$ give different contributions in the amplitude 
of monopole and quadrupole spectra, and their total contribution can reach
$\sim10\%$ and $\sim40\%$ for monopole and quadrupole spectra at 
$k\lesssim0.2h$Mpc$^{-1}$, 
respectively. Thus, even though the resultant shape of the total spectrum  
$P^{\rm(S)}(k)$ apparently resembles the one obtained from phenomenological 
model, the actual contribution of the corrections $A$ and $B$ would be 
large and cannot be neglected.

Note, however, that a closer look at low-$z$ behavior reveals a slight 
discrepancy around $k\sim0.15h$Mpc$^{-1}$ and $0.22h$Mpc$^{-1}$  
in the monopole spectrum. Also, discrepancies in the quadrupole spectrum 
seems bit large,  
and eventually reach $\sim5\%$ error in some wavenumbers at $z=0.5$. 
This is partially ascribed to our heterogeneous treatment 
on the corrections $A$ and $B$ using the standard PT calculations. 
It is known that the standard PT result generically gives rise to a 
strong damping in the BAOs, and it incorrectly leads to a phase reversal 
of the BAOs. Thus, beyond the validity regime 
of the standard PT, the predictions including the small corrections 
tend to oversmear the acoustic feature, leading to 
a small discrepancy shown in Fig.~\ref{fig:ratio_pk_red2}.
\begin{figure}[t]

\vspace*{-0.5cm}

\begin{center}
\includegraphics[width=8.5cm,angle=0]{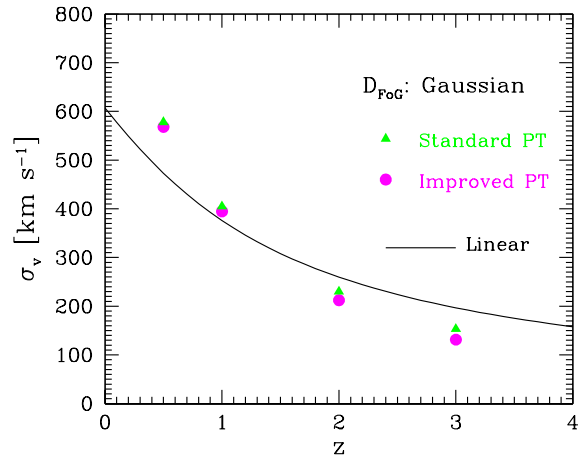}
\end{center}

\vspace*{-1.5cm}

\caption{Same as in Fig.~\ref{fig:sigma_v}, but we here adopt new model 
  of redshift distortion in estimating $\sigma_{\rm v}$. The filled 
  triangle and circles are the results obtained from predictions based on 
  standard PT and improved PT calculations, respectively (see dashed and 
  solid lines in Fig.~\ref{fig:ratio_pk_red2}. 
\label{fig:sigma_v2}}
\end{figure}

Another source for the discrepancies may come from 
the effect of finite-mode sampling caused by the finite boxsize of 
the N-body simulations.  
As advocated by Refs.\cite{Takahashi:2008wn, Nishimichi:2008ry}, 
due to the finite number of Fourier modes, 
the matter power spectrum measured from N-body simulations may not agree 
well with the predictions of linear theory nor standard PT even at very 
large scales, and tends to systematically deviate from them. 
While we follow and extend the procedure of Ref.\cite{Nishimichi:2008ry} 
to correct this effect in redshift space, it relies on the leading-order 
calculations of standard PT, and the correction for finite-mode 
sampling has been restricted to the low-$k$ modes, $k\lesssim0.1h$Mpc$^{-1}$ 
\cite{Taruya:2009ir}. Hence, the 
high-$k$ modes of the power spectrum plotted here 
may be affected by the effect of finite-mode sampling, and    
it would be significant for 
higher-multipole spectrum because of its small amplitude.  
This might be still serious even 
with the 30 independent data of N-body simulations.

Perhaps, the best way to remedy these discrepancies at low-$z$ 
is both to apply the improved PT treatment to the corrections $A$ and $B$, 
and to consider the higher-order contributions for correcting the effect of 
finite-mode sampling over the relevant range of BAOs.
The complete analysis along the line of 
this need some progress and is beyond the scope of this paper. 
Nevertheless, it should be stressed that the model given by 
Eq. (\ref{eq:new_model})  captures several important 
aspects of redshift distortion, and even the 
present treatment with standard PT calculations of the 
corrections $A$ and $B$ can provide a better description for 
power spectra. In Fig.~\ref{fig:sigma_v2}, 
we plot the fitted values of the velocity dispersion obtained 
from the new predictions shown in Fig.~\ref{fig:ratio_pk_red2}. 
The redshift dependence of the fitted results roughly matches 
physical intuition, and is rather consistent with the 
linear theory prediction. This is contrasted to 
the cases neglecting the corrections 
(see Fig.~\ref{fig:sigma_v}).

\begin{figure}[t]

\vspace*{-1.0cm}

\begin{center}
\includegraphics[width=9cm,angle=0]{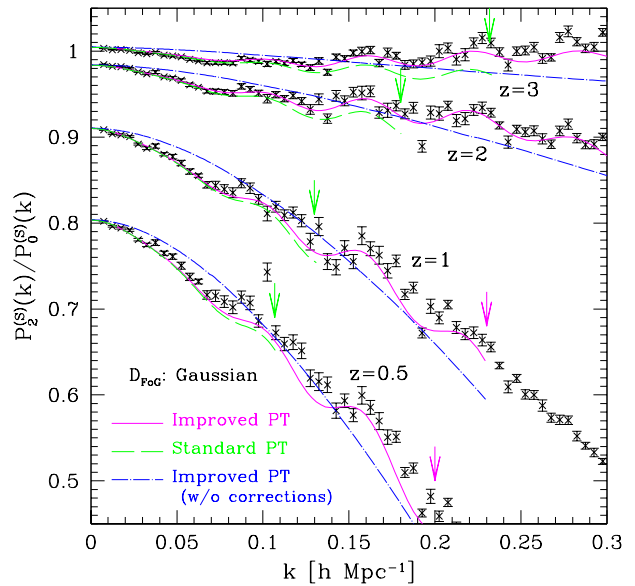}
\end{center}

\vspace*{-0.8cm}

\caption{Quadrupole-to-monopole ratios for redshift-space 
  power spectrum, $P^{\rm(S)}_2(k)/P^{\rm(S)}_0(k)$, given at $z=3$, $2$, 
  $1$, and $0.5$ (from top to bottom). Solid and dashed lines respectively 
  represent the predictions based on new model of redshift 
  distortion combining improved PT and standard PT calculation to 
  estimate the three different power spectra $\Pdd$, $\Pdv$ and $\Pvv$. 
  Dot-dashed lines are the results based on the phenomenological model 
  neglecting the corrections, which correspond to 
  solid lines in Fig.~\ref{fig:ratio_pk_red_phenom} (i.e., 
  non-linear $\PKaiser$ $+$ Gaussian $\DFoG$). 
  The vertical arrows indicate the maximum wavenumber $k_{1\%}$ 
  for standard PT (left) and improved PT (right). 
\label{fig:qm_ratio}}
\end{figure}

As another significance, 
we plot in Fig.~\ref{fig:qm_ratio} the quadrupole-to-monopole ratios  
for redshift-space power spectra. The new model predictions using 
standard and improved PT calculations (solid and dashed) 
are compared with those neglecting 
the corrections $A$ and $B$ (dot-dashed). The amplitude of the 
ratio $P_2^{\rm(S)}/P_0^{\rm(S)}$ basically reflects the 
strength of the clustering anisotropies, and is proportional to 
$(4f/3+4f^2/7)/(1+2f/3+f^2/5)$ in the limit $k\to0$ 
(e.g., \cite{Kaiser:1987qv,Hamilton:1997zq,1992ApJ385L5H}). 
One noticeable point is that the N-body results of quadrupole-to-monopole 
ratio do exhibit an oscillatory behavior, and the model including 
the corrections (\ref{eq:new_model}) reproduces the N-body trends 
fairly well. On the other hand, the phenomenological model neglecting 
the corrections generally predicts the smooth scale-dependence of 
the ratio $P_2^{\rm(S)}/P_0^{\rm(S)}$, 
and thus it fails to reproduce the oscillatory feature. 
Since this oscillation is originated from the acoustic feature in BAOs, 
Fig.~\ref{fig:qm_ratio} implies that the quadrupole-to-monopole ratio 
possesses helpful information not only to constrain the growth-rate 
parameter $f$, but also to determine the acoustic scales. In other words, 
any theoretical template for redshift-space power spectrum neglecting 
the corrections $A$ and $B$ may produce a systematic bias in 
determining the growth-rate parameter $f(z)$, Hubble parameter $H(z)$ 
and angular diameter distance $D_A(z)$, which we will discuss in details 
in next section. 

\section{Implications} 
\label{sec:implication}

The primary science goal of future galaxy surveys is to clarify 
the nature of late-time cosmic acceleration, and thereby
constraining the parameters $D_A(z)$, $H(z)$ and $f(z)$ through 
a precise measurement of BAOs in redshift space would be the most
important task. 
However, these constraints may be biased if we use the 
incorrect model of redshift distortion as theoretical template 
fitting to observations. In this section, we explore the potential 
impact on the uncertainty and bias in the parameter estimation for 
$D_A(z)$, $H(z)$ and $f(z)$.

\subsection{Recovery of parameters $D_A$, $H$ and $f$}
\label{subsec:recovery_DA_H_f}

Let us first examine the parameter estimation 
using the new model of redshift distortion. Fitting 
the theoretical template of power spectrum to the N-body data, 
we will check if the best-fit parameters for 
$D_A(z)$, $H(z)$ and $f(z)$ can be correctly recovered from 
the monopole and quadrupole moments of anisotropic BAOs.

We model the power spectrum of N-body simulations by 
\begin{equation}
P_{\rm model}^{\rm(S)}(k,\mu) = \frac{H(z)}{H_{\rm fid}(z)}
\left\{ \frac{D_{A,{\rm fid}}(z)}{D_A(z)}\right\}^2\,\,
P^{\rm(S)}(q,\nu), 
\label{eq:pk_model} 
\end{equation}
where the comoving wavenumber $k$ and the directional cosine $\mu$ for 
the underlying cosmological model are 
related to the true ones $q$ and $\nu$ by the Alcock-Paczynski effect 
through (e.g., \cite{Ballinger:1996cd,Magira:1999bn,Padmanabhan:2008ag})
\begin{align}
&q=k\,\left[
\left(\frac{D_{A,{\rm fid}}}{D_A}\right)^2+
\left\{\left(\frac{H}{H_{\rm fid}}\right)-
\left(\frac{D_{A,{\rm fid}}}{D_A}\right)^2\right\}\mu^2
\right]^{1/2},
\\
&\nu= \left(\frac{H}{H_{\rm fid}}\right)\,\mu
\nonumber\\
&\qquad\,\times\left[
\left(\frac{D_{A,{\rm fid}}}{D_A}\right)^2+
\left\{\left(\frac{H}{H_{\rm fid}}\right)-
\left(\frac{D_{A,{\rm fid}}}{D_A}\right)^2\right\}\mu^2
\right]^{-1/2}, 
\end{align}
The quantities $D_{A,{\rm fid}}$ and $H_{\rm fid}$ are the fiducial 
values of the angular diameter distance and Hubble parameter adopted in 
the N-body simulations. For a given set of cosmological parameters, 
the redshift-space power spectrum $P^{\rm(S)}$ is calculated from 
Eq.~(\ref{eq:new_model}), but we here treat the quantity $f$ as free 
parameter in addition to the velocity dispersion $\sigmav$. Further, 
to mimic a practical data analysis using galaxy power spectrum, 
we introduce the bias parameter $b$, assuming the 
linear deterministic relation, i.e., $\delta_{\rm sim}=b\,\delta_{\rm m}$ 
\footnote{In the case adopting linear galaxy bias, 
the growth rate parameter $f$ and the power spectra $P_{ab}$ 
in the expression (\ref{eq:new_model}) 
are respectively replaced with $\beta\equiv f/b$ and $b^2P_{ab}$. 
Also, the standard PT expression for the corrections 
$A(k,\mu;f)$ and $B(k,\mu;f)$ should be replaced with 
$b^3\,A(k,\mu;\beta)$ and  $b^4\,B(k,\mu;\beta)$. }. 
Then, fitting the monopole and quadrupole power spectra of 
Eq.~(\ref{eq:pk_model}) to those of the N-body simulation at $z=1$,  
we determine the best-fit values of $D_A$, $H$ and $f$, 
just marginalized over the parameters $\sigmav$ and $b$. To do this, we use 
the Markov chain Monte Carlo (MCMC) technique described by 
Ref.~\cite{Lewis:2002ah}, and adopt the likelihood 
function given by
\begin{align}
&-2\ln\mathcal{L}=\sum_{n}\sum_{\ell,\ell'=0,2}
\left\{P_{\ell,{\rm sim}}^{\rm(S)}(k_n)-
P_{\ell,{\rm model}}^{\rm(S)}(k_n)\right\}
\nonumber\\
&\qquad\qquad\times
\mbox{Cov}_{\ell,\ell'}^{-1}(k_n)\,
\left\{P_{\ell',{\rm sim}}^{\rm(S)}(k_n)-
P_{\ell',{\rm model}}^{\rm(S)}(k_n)\right\}, 
\label{eq:likelihood_func}
\end{align}
where the quantity $\mbox{Cov}_{\ell,\ell'}$ represents the covariance 
matrix between different multipoles. The range of wavenumber used in 
the likelihood analysis was chosen as $k\leq k_{\rm max}=0.205h$Mpc$^{-1}$, 
so as to satisfy $k_{\rm max}\leq k_{1\%}$. 
As for the covariance, we simply  
ignore the non-Gaussian contribution (see Ref.~\cite{Takahashi:2009bq} 
for validity of this treatment), 
and use the linear theory to estimate the diagonal components of the 
covariance, 
$\mbox{Cov}_{\ell,\ell'}$, including the effect of shot-noise contribution 
assuming the galaxy number density 
$\overline{n}_{\rm g}=5\times10^{-4}h^3$Mpc$^{-3}$. The explicit expression 
for the covariance is presented in Appendix \ref{appendix:covariance}. 
We checked that the linear theory estimate 
reasonably reproduces the N-body results of the covariance matrix 
for the range of our interest $k\lesssim0.3h$Mpc$^{-1}$ at $z=1$.

\begin{figure}[t]
\hspace*{-0.5cm}
\includegraphics[width=10cm]{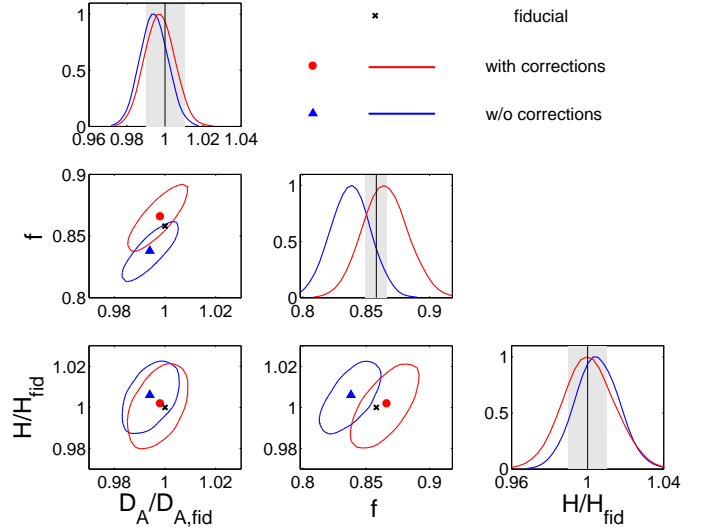}

\vspace*{0.0cm}

\caption{Results of MCMC analysis using the model of redshift distortion 
with and without corrections (depicted as blue and red lines, 
respectively). 
Based on the power spectrum template (\ref{eq:pk_model}), we derive 
the posterior distribution for the parameters $D_A$, $H$ and $f$ from 
the monopole and quadrupole spectra of N-body simulations at $z=1$, 
marginalized over the one-dimensional velocity dispersion $\sigmav$ and 
linear bias parameter $b$. Top left, 
middle center and bottom right show the marginalized posterior distribution  
for $D_A/D_{A,{\rm fid}}$, $H/H_{\rm fid}$ and $f$. Shaded regions indicate 
the $1\%$ interval around the fiducial values. 
Middle left, bottom left, 
and bottom center plot the two-dimensional $1\mbox{-}\sigma$ errors on 
the surfaces $(H/H_{\rm fid}, f)$, $(D_A/D_{A,{\rm fid}}, H/H_{\rm fid})$, and 
$(f, H/H_{\rm fid})$. Note that in estimating likelihood function 
(\ref{eq:likelihood_func}), 
we adopted the linear theory to calculate the covariance matrix 
$\mbox{Cov}_{\ell,\ell'}$, including the shot-noise contribution with 
$\overline{n}_{\rm g}=5\times10^{-4}h^3$Mpc$^{-3}$ and assuming an
idealistically large survey volume $V_s=20h^{-3}$Mpc$^3$ 
(see Appendix \ref{appendix:covariance} for explicit expression). 
\label{fig:MCMC}}
\end{figure}

Fig.~\ref{fig:MCMC} summarizes the result of the MCMC analysis assuming 
an idealistically large survey with $V_s=20h^{-3}$Gpc$^3$. The 
two-dimensional contour of the $1\mbox{-}\sigma$ marginalized errors 
are shown for $D_A/D_{A,{\rm fid}}$ vs $H/H_{\rm fid}$ (bottom left),  
$D_A/D_{A,{\rm fid}}$ vs $f$ (middle left),  and $f$ vs $D_A/D_{A,{\rm fid}}$ 
(bottom center). Also, the marginalized posterior distribution for each 
parameter are plotted in the top left, middle center, and bottom right panels. 
In each panel, blue and red lines respectively represent the 
results using the model of redshift distortion with and without
the terms $A$ and $B$.

As it is clear from Fig.~\ref{fig:MCMC},  
the model including the corrections shows a better performance.   
Within the $1\mbox{-}\sigma$ errors, which  
roughly correspond to the precision of a percent-level, 
it correctly reproduces the fiducial values of the parameters 
(indicated by crosses). On the other hand, the two-dimensional errors of 
the results neglecting the corrections show a clear evidence for 
the systematic bias on the best-fit parameters. Accordingly, 
the resultant value of $\chi^2$ around the best-fit parameters,  
given by $\chi^2=-2\ln\mathcal{L}$, is larger than that 
of the case including the corrections: $\chi^2=10.1$ and $22.2$ for 
the cases with and without corrections, respectively. Although the 
deviation from the fiducial values seems somewhat small except for the 
growth-rate parameter $f$, this is solely due to the fact that 
we only use the monopole and quadrupole power spectra. It 
would be generally significant in the analysis using the full shape of 
redshift-space power spectrum, for which 
the statistical errors are greatly reduced, and thereby 
the systematic biases would be prominent.

\begin{figure*}[t]
\begin{center}
\includegraphics[width=8.5cm,angle=0]{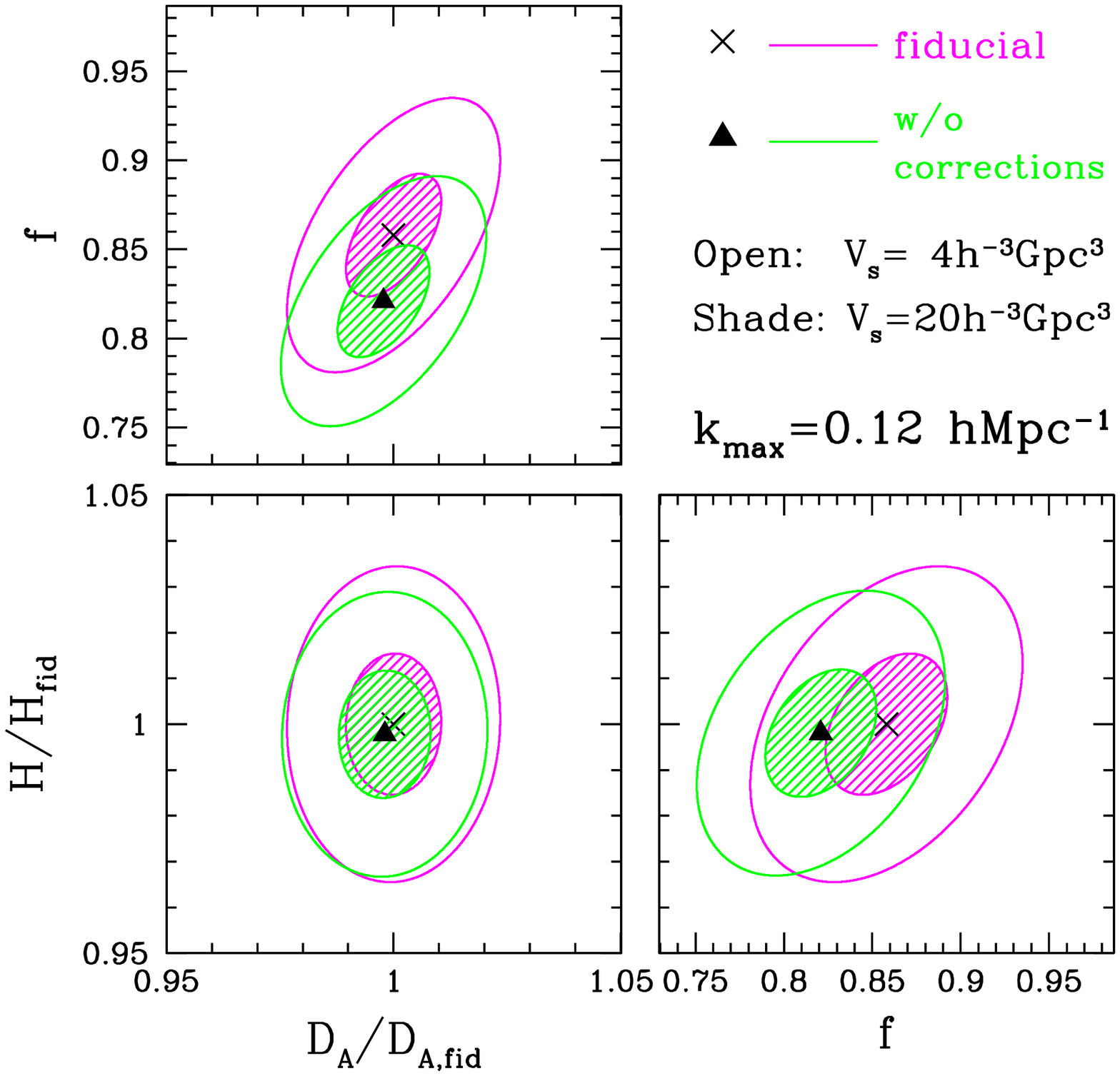}
\includegraphics[width=8.5cm,angle=0]{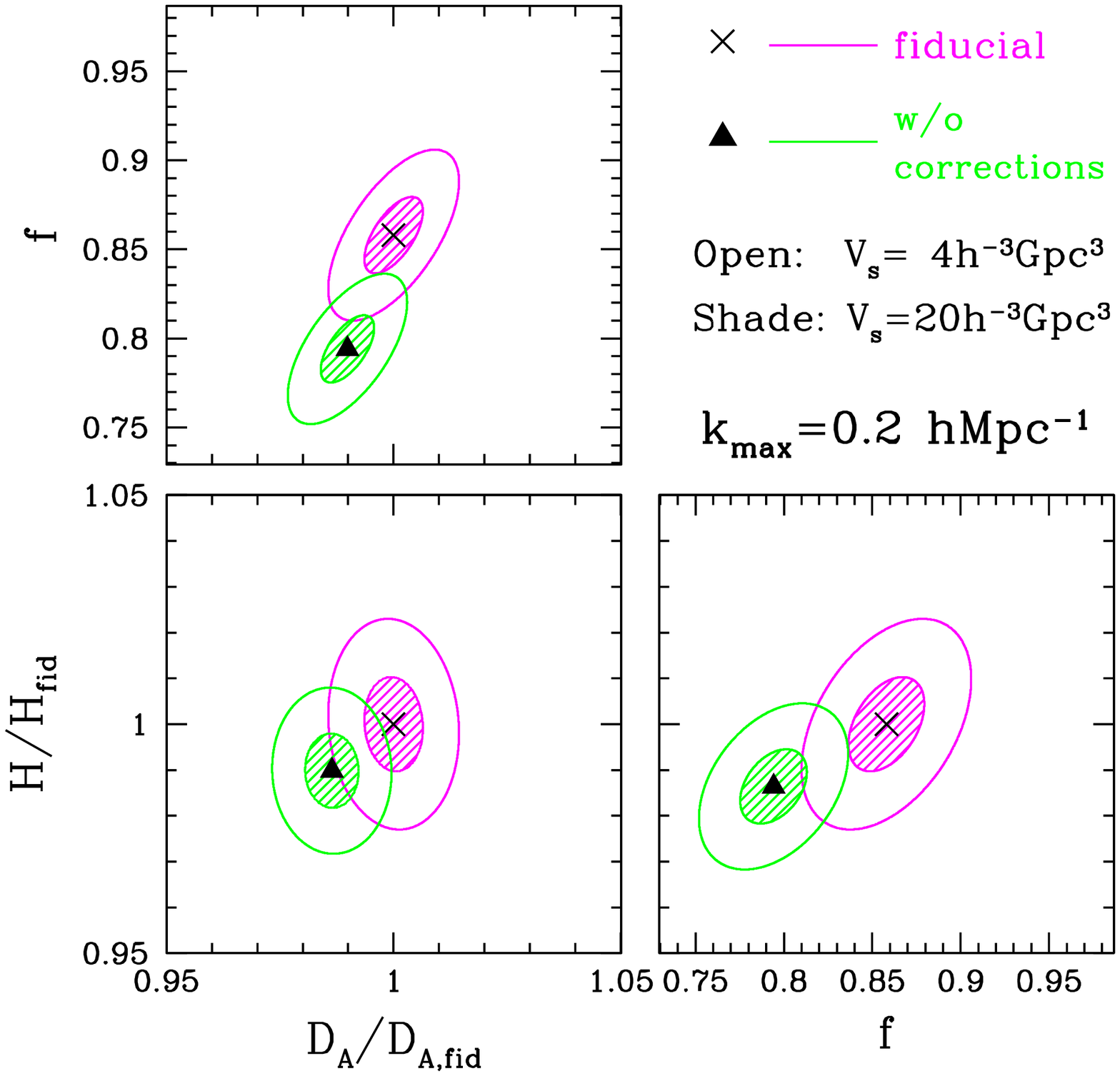}
\end{center}

\vspace*{-0.8cm}

\caption{Expected two-dimensional contours on marginalized errors 
around the best-fit values 
of $D_A/D_{A,{\rm fid}}$ vs $H/H_{\rm fid}$ (bottom left), 
$f$ vs $H/H_{\rm fid}$ (bottom right) and  
$D_A/D_{A,{\rm fid}}$ vs $f$ (top left) at $z=1$, obtained from 
the {\it full shape} of redshift-space power spectrum. 
The maximum wavenumber for parameter estimation is chosen as 
$k_{\rm max}=0.12$ (left) and $0.2h$Mpc$^{-1}$ (right), so as to
satisfy the condition $k_{\rm max}<k_{1\%}$ for 
standard PT and improved PT, respectively. 
In each panel, open and shaded contours indicate the 
two dimensional errors for the surveys with $V_s=4$ and $20h^{-3}$Gpc$^3$.  
\label{fig:error_f_DA_H}}
\end{figure*}

\subsection{Impact of redshift distortion on 
future measurements of $D_A$, $H$ and $f$} 
\label{subsec:impact_on_H_DA}

Given the fact that the robust measurement of $D_A$, $H$ and $f$ can be 
made with the new model of redshift distortion,  
we then move to the discussion on the potential impact on 
the future measurements using the {\it full shape} of the redshift-space 
power spectrum. Here, for illustrative purpose, we 
consider the two surveys around $z=1$, with volume 
$V_s=4$ and $20\,h^{-3}$Gpc$^3$, and quantitatively 
estimate how the wrong model of 
redshift distortion leads to the incorrect measurements of 
$D_A$, $H$ and $f$.

The fundamental basis to estimate 
the uncertainties and systematic biases on model parameters is 
the Fisher matrix formalism. The Fisher matrix for galaxy survey is given by 
\begin{align}
&F_{ij}=\frac{V_s}{(2\pi)^2}\int_{k_{\rm min}}^{k_{\rm max}} 
dk\,k^2\,\int_{-1}^1d\mu
\nonumber\\
&\quad\times\,\frac{\partial \ln P_{\rm obs}^{\rm(S)}(k,\mu)}{\partial p_i}
\frac{\partial \ln P_{\rm obs}^{\rm(S)}(k,\mu)}{\partial p_j}
\left\{\frac{\overline{n}_{\rm g}\,P_{\rm obs}^{\rm(S)}(k,\mu)}
{\overline{n}_{\rm g}\,P_{\rm obs}^{\rm(S)}(k,\mu)+1}\right\}^2
\label{eq:Fisher_matrix}
\end{align}
with $\overline{n}_{\rm gal}$ being the number density of galaxies, for 
which we specifically set 
$\overline{n}_{\rm gal}=5\times10^{-4}\,h^3$Mpc$^{-3}$. 
The minimum wavenumber available for a given survey, $k_{\rm min}$,  
is set to $2\pi/V_s^{1/3}$. 
Here, the observed power spectrum $P_{\rm obs}^{\rm(S)}$ is given 
by Eq.~(\ref{eq:pk_model}), and we allow to include the influence of 
galaxy biasing adopting the deterministic linear relation, 
$\delta_{\rm gal}=b\,\delta_{\rm m}$. Then, 
we have five parameters in total, given by 
$p_i=\{b,\,f,\,\sigmav,\,D_A/D_{A,{\rm fid}},\,H/H_{\rm fid} \}$.  
Fiducial values of these parameters 
are set as $b=2$, $f=0.858$, $D_A/D_{A,{\rm fid}}=1$ 
and $H/H_{\rm fid}=1$. As for the velocity dispersion $\sigmav$, we use 
the fitted result to N-body simulations adopting the new model of 
redshift distortion in Sec.~\ref{subsec:comparison}, 
and set $\sigmav=395$\,km\,s$^{-1}$.

Based on the Fisher matrix (\ref{eq:Fisher_matrix}), the systematic 
bias for parameter $p_i$ caused by incorrectly modeling 
theoretical power spectrum is estimated from the following formula: 
\begin{equation}
\Delta p_i = -\sum_j \left(F'^{-1}\right)_{ij}\,s_j
\label{eq:systematic_bias}
\end{equation}
where $F'^{-1}$ is the inverse Fisher matrix evaluated at the fiducial 
parameter set, but is obtained from an incorrect model of 
redshift distortion as a theoretical template of redshift-space power spectrum. 
The vector $s_j$ is given by 
\begin{align}
&s_j=\frac{V_s}{(2\pi)^2}\int dk\,k^2\,\int_{-1}^1d\mu\,
\frac{P^{\rm sys}(k,\mu)}{P^{\rm wrong}(k,\mu)} 
\frac{\partial \ln P^{\rm wrong}(k,\mu)}{\partial p_j} 
\nonumber\\
&\qquad\qquad\qquad
\times\left\{\frac{\overline{n}_{\rm g}\,P^{\rm wrong}(k,\mu)}
{1+\overline{n}_{\rm g}\,P^{\rm wrong}(k,\mu)}\right\}^2.
\end{align}
The function $P^{\rm wrong}(k,\mu)$ is the theoretical template
adopting the incorrect model of redshift distortion. The 
systematic differences in the power spectrum amplitude are 
quantified as $P^{\rm sys}(k,\mu)=P^{\rm wrong}(k,\mu)-P^{\rm true}(k,\mu)$, 
where $P^{\rm true}(k,\mu)$ 
is the correct template for redshift-space power spectrum 
$P_{\rm obs}^{\rm(S)}$, for which we 
assume the new model of redshift distortion including the 
terms $A$ and $B$ (Eq.(\ref{eq:new_model})). Below, we 
will quantify the magnitude of systematic biases if we 
incorrectly apply the model of redshift distortion neglecting the corrections 
$A$ and $B$ for the power spectrum template.

Fig.~\ref{fig:error_f_DA_H} plots the results of the Fisher matrix 
calculations marginalized over the nuisance parameters $b$ and $\sigmav$. 
The uncertainties and biases for the best-fit values of $f$, $D_A$ and $H$ 
are estimated assuming $k_{\rm max}=0.12h$Mpc$^{-1}$ (left) and 
$0.2h$Mpc$^{-1}$ (right), and the results are shown for the surveys with 
$V_s=4\,h^{-3}$Gpc$^3$ (open) and $20\,h^{-3}$Gpc$^3$ (shade). 
In each panel, two-dimensional contours around the crosses and filled 
triangles show the expected 1-$\sigma$ ($68\%$ C.L.) errors around the 
best-fit values adopting the model of redshift distortion with and without 
the corrections, respectively. 
The differences between best-fit values (crosses and filled triangles) 
represent the systematic biases estimated from 
Eq.~(\ref{eq:systematic_bias}), which remain unchanged irrespective 
of the survey volume $V_s$. Since 
the size of marginalized uncertainties is proportional to $V_s^{-1/2}$, 
the systematic bias in the best-fit parameters become relatively prominent
and is considered to be a serious problem if we increase the survey volume. 
Note that similar to the result in Fig.~\ref{fig:MCMC}, 
there exists a tight correlation of the parameters between 
the growth rate parameter $f$ and quantities $D_A$ and $H$. This 
is consistent with the finding by Ref.~\cite{Simpson:2009zj}, 
indicating that the distinguishing dark energy from modified gravity 
needs another observational constraint.

Fig.~\ref{fig:error_f_DA_H} implies
that phenomenological model of redshift distortion 
neglecting the corrections can produce a large systematic error. 
As increasing the maximum wavenumber $k_{\rm max}$, the bias on the 
measurements of angular diameter distance and Hubble parameter 
reaches $\sim1-2\%$ error, while 
the best-fit value for the growth rate parameter 
would be seriously biased with $\sim5\%$ error. 
If we conservatively choose a smaller value of 
$k_{\rm max}\lesssim0.12h$Mpc$^{-1}$,   
these systematics could be still within the size of statistical error 
for surveys with typical volume of $V_s\sim4\,h^{-3}$Gpc$^3$. 
However, for a survey of larger volume with $V_s\gtrsim20\,h^{-3}$Gpc$^3$, 
the systematic error on the growth rate parameter becomes 
outside the the marginalized uncertainty. If we aggressively 
choose $k_{\rm max}\sim0.2h$Mpc$^{-1}$ in order to 
reduce statistical uncertainties, the systematic biases 
become definitely serious issues in all of the parameters $f$, $D_A$ 
and $H$ for both surveys of volume $V_s=4$ and $20h^{-3}$Gpc$^3$. 
Hence, correctly modeling redshift distortion would be very 
crucial for both stage-III and -IV class surveys 
defined by the Dark Energy Task Force \cite{Albrecht:2006um}.

\section{Discussion and conclusion}
\label{sec:conclusion}

In this paper, 
we have investigated the power spectrum in redshift space, and presented 
a new model of redshift distortion, 
which is particularly suited for modeling anisotropic 
BAOs around $k=0\sim0.3\,h$Mpc$^{-1}$. Contrary to the 
previous phenomenological modes in which the effects of Kaiser and 
Finger-of-God are separately treated in a multiplicative way, 
the new model includes the corrections coming from 
the non-linear coupling between velocity and density fields, 
which give rise to a slight uplift in the amplitude of 
monopole and quadrupole 
power spectra. The model predictions can give a good agreement with 
results of N-body simulations, and a percent level precision is almost 
achieved.

Based on the new model of redshift distortion, we proceeded to 
the parameter estimation analysis, and checked if the theoretical 
prediction correctly
recovers the cosmological information from the 
monopole and quadrupole spectra of N-body simulations. 
MCMC analysis revealed that while the 
new model of redshift distortion combining the improved PT calculation 
faithfully reproduces the fiducial parameters $D_A$, $H$ and $f$ 
and the precision can reach at a percent level,  
the model neglecting the corrections ($A$ and $B$ terms) exhibits 
a slight offset of the best-fit values. In order to 
estimate the potential impact on the future measurement, 
we have further made the Fisher matrix analysis 
using the full shape of power spectrum $P^{\rm(S)}(k,\mu)$, and 
found that the existing phenomenological models of redshift distortion 
neglecting the corrections produce a systematic error on measurements of 
the angular diameter distance and Hubble parameter by $1\sim2\%$,  
and the growth rate parameter by $\sim5\%$. This would become 
non-negligible for stage-III and -IV class surveys 
defined by the Dark Energy Task Force. Correctly modeling 
redshift distortion is thus crucial, and the new prescription of 
redshift-space power spectrum presented here plays an 
essential role in constraining the dark energy and/or modified gravity 
from anisotropic BAOs.

Finally, we note several remaining tasks 
in practical application to the precision measurement of 
BAOs. One is the improved treatment for calculation of the corrections, 
$A$ and $B$ terms, which needs to evaluate the bispectrum of density 
and velocity fields. In doing this, a systematic treatment using 
multi-point propagator developed by Ref.~\cite{Bernardeau:2008fa} would 
be useful and indispensable. Also, the effects of 
the new contributions to the redshift-space clustering in the presence 
of the primordial non-Gaussianity and the dark sector 
interaction would be presumably important 
(e.g., \cite{Schmidt:2010pf,Koyama:2009gd,Simpson:2010yt}), 
and should deserve further investigation. Of course, the biggest issue is 
the galaxy biasing. Recent numerical and analytical studies 
claim that the scale-dependent and stochastic properties of the galaxy bias 
can change the redshift-space power spectrum, and the potential impact 
on the determination of the growth-rate parameter would be significant 
\cite{Desjacques:2009kt,Okumura:2010sv}. A realistic modeling of galaxy 
biasing relevant for the scale of BAOs is thus essential, and a further 
improvement of the power spectrum template needs to be developed.

\begin{acknowledgements}
We would like to thank Yasushi Suto and Kazuhiro Yamamoto for comments 
and discussion.  
AT is supported by a Grant-in-Aid for Scientific 
Research from the Japan Society for the Promotion of Science (JSPS) 
(No.~21740168). TN and SS are supported from JSPS. 
This work was supported in part by 
Grant-in-Aid for Scientific Research on Priority Areas No.~467 
``Probing the Dark Energy through an Extremely Wide and Deep Survey with 
Subaru Telescope'', and JSPS Core-to-Core Program ``International 
Research Network for Dark Energy''.
\end{acknowledgements}

\bigskip

\appendix
\section{Perturbation theory calculations for correction terms}
\label{appendix:PT_calc_correction}

In this Appendix, we present the perturbative expressions for the 
corrections $A$ and $B$ defined in Eqs.~(\ref{eq:A_term}) and 
(\ref{eq:B_term}), which are originated from the coupling between 
Kaiser and Finger-of-God effects.

Let us first consider the correction $A$, which involves the bispectrum 
$B_\sigma$ of density and velocity divergence 
(see Eq.~(\ref{eq:def_B_sigma})). Using the perturbative 
solutions up to the second-order, the leading-order result of the 
bispectrum becomes   
\begin{widetext}
 \begin{align}
& B_\sigma(\bfk_1,\bfk_2,\bfk_3)=(-2f)
 \left[\left(1+\frac{k_{2z}^2}{k_2^2}f\right)
\left(1+\frac{k_{3z}^2}{k_3^2}f\right)
 \,G_2(\bfk_2,\bfk_3)\,\Plin(k_2)\Plin(P_3)\right.
\nonumber\\
&\quad\quad\quad+\left(1+\frac{k_{3z}^2}{k_3^2}f\right)
\left\{F_2(\bfk_1,\bfk_3)+\frac{k_{2z}^2}{k_2^2}f\, 
G_2(\bfk_1,\bfk_3)\,\right\}\Plin(k_1)\Plin(P_3)
\nonumber\\
&\quad\quad\quad\left.+\left(1+\frac{k_{2z}^2}{k_2^2}f\right)
\left\{F_2(\bfk_1,\bfk_2)+\frac{k_{3z}^2}{k_3^2}f\, 
G_2(\bfk_1,\bfk_2)\,\right\}\Plin(k_1)\Plin(P_2)\right]
\label{eq:bispectrum_PT}
 \end{align}
\end{widetext}
with $F_2$ and $G_2$ being the second-order perturbation kernels 
given by (e.g., \cite{Bernardeau:2001qr,Crocce:2005xy,Nishimichi:2007xt})
\begin{align}
&F_2(\bfk_1,\bfk_2)=
\frac{5}{7}+\frac{\bfk_1\cdot\bfk_2}{2k_1k_2}\left(\frac{k_1}{k_2}+
\frac{k_2}{k_1}\right)+\frac{2}{7}\left(\frac{\bfk_1\cdot\bfk_2}{k_1k_2}
\right)^2,
\nonumber\\
&G_2(\bfk_1,\bfk_2)=
\frac{3}{7}+\frac{\bfk_1\cdot\bfk_2}{2k_1k_2}\left(\frac{k_1}{k_2}+
\frac{k_2}{k_1}\right)+\frac{4}{7}\left(\frac{\bfk_1\cdot\bfk_2}{k_1k_2}
\right)^2.
\nonumber
\end{align}
Note that the bispectrum (\ref{eq:bispectrum_PT}) possesses the 
following symmetries: 
$B_\sigma(\bfk_1,\bfk_2,\bfk_3)=B_\sigma(\bfk_1,\bfk_3,\bfk_2)=
B_\sigma(-\bfk_1,-\bfk_2,-\bfk_3)$. 
Then, substituting the expression (\ref{eq:bispectrum_PT}) into 
the definition (\ref{eq:A_term}), the correction $A$ can be recast 
schematically in the form as
\begin{align}
A(k,\mu)=-k\mu\,\sum_{m,n}   \int\frac{d^3\bfp}{(2\pi)^3}\,
f^mp_z^n\,\,\,Q_{mn}(\bfk,\bfp), 
\end{align}
where the function $Q_{mn}$ is the scalar function of $\bfk$ and $\bfp$. 
To further perform the angular integral, we use the formulae presented in 
Appendix \ref{appendix:formula} (Eq.~(\ref{eq:integration_formula})), 
which can be obtained by utilizing the 
rotational covariance of the integral. After straightforward 
but lengthy calculation, 
the correction $A(k,\mu)$ is finally reduced to the following form: 
\begin{widetext}
\begin{align}
&A(k,\mu;z)=
\sum_{m,n=1}^3\, \mu^{2m}\,f^{n} 
\frac{k^3}{(2\pi)^2}
\left[\int_0^\infty dr \int_{-1}^{+1} dx \,\,\Bigl\{
\,A_{mn}(r,x)\,\Plin(k;z) 
+ \widetilde{A}_{mn}(r,x)\,\Plin(kr;z) \,
\Bigr\}
\right.
\nonumber\\
&\qquad\qquad\qquad\qquad\qquad\qquad\qquad\qquad\left.
\times \frac{\Plin\left(k\sqrt{1+r^2-2rx};z\right)}{(1+r^2-2rx)^2}
+\Plin(k;z) \,\int_0^\infty dr \,a_{mn}(r)\,\Plin(kr;z) \right], 
\label{eq:A_PT_formula}
\end{align}
\end{widetext}
where we introduce the quantities 
$r=k/p$ and $x=(\bfk\cdot\bfp)/k/p$. Note again that 
$\mu$ is the cosine of the angle between line-of-sight direction 
$\widehat{z}$ and the vector $\bfk$, i.e., $\mu=(\bfk\cdot\widehat{z})/k$. 
The non-vanishing components of $A_{mn}$, $\widetilde{A}_{mn}$ and 
$a_{mn}$ are 
\begin{align}
A_{11}=&-\frac{r^3}{7}\left\{x+6x^3+r^2x(-3+10x^2)\right.
\nonumber\\
&\left.
+r\,(-3+x^2-12x^4)\right\},
\nonumber
\end{align}
\begin{align}
 A_{12}=\frac{r^4}{14}(x^2-1)(-1+7rx-6x^2),
\nonumber
\end{align}
\begin{align}
 A_{22}=&\frac{r^3}{14}\left\{r^2x(13-41x^2)-4(x+6x^3)
\right.
\nonumber\\
&\left.
+r\,(5+9x^2+42x^4)\right\},
\nonumber
\end{align}
\begin{align}
A_{23}=A_{12},
\nonumber
\end{align}
\begin{align}
A_{33}=\frac{r^3}{14}(1-7rx+6x^2)\left\{-2x+r(-1+3x^2)
\right\},
\nonumber
\end{align}
for $A_{mn}$, 
\begin{align}
\widetilde{A}_{11}=\frac{1}{7}(x+r-2rx^2)(3r+7x-10rx^2),
\nonumber
\end{align}
\begin{align}
\widetilde{A}_{12}=\frac{r}{14}(x^2-1)(3r+7x-10rx^2),
\nonumber
\end{align}
\begin{align}
\widetilde{A}_{22}=\frac{1}{14}
\left\{28x^2+rx(25-81x^2)+r^2(1-27x^2+54x^4)\right\},
\nonumber
\end{align}
\begin{align}
\widetilde{A}_{23}=\frac{r}{14}(1-x^2)(r-7x+6rx^2),
\nonumber
\end{align}
\begin{align}
\widetilde{A}_{33}=\frac{1}{14}(r-7x+6rx^2)(-2x-r+3rx^2),
\nonumber
\end{align}
for $\widetilde{A}_{mn}$, and
\begin{align}
a_{11}=&-\frac{1}{84r}\left[2r(19-24r^2+9r^4)\right.
\nonumber\\
&\left.\qquad
-9(r^2-1)^3\log\left|\frac{r+1}{r-1}\right|\right],
\nonumber
\end{align}
\begin{align}
a_{12}=&\frac{1}{112r^3}\left[2r(r^2+1)(3-14r^2+3r^4)\right.
\nonumber\\
&\left.\qquad
-3(r^2-1)^4\log\left|\frac{r+1}{r-1}\right|\right],
\nonumber
\end{align}
\begin{align}
a_{22}=&\frac{1}{336r^3}\left[2r(9-185r^2+159r^4-63r^6)\right.
\nonumber\\
&\left.\qquad
+9(r^2-1)^3(7r^2+1)\log\left|\frac{r+1}{r-1}\right|\right],
\nonumber
\end{align}
\begin{align}
& a_{23}=a_{12},
\nonumber
\end{align}
\begin{align}
 a_{33}=&\frac{1}{336r^3}\left[2r(9-109r^2+63r^4-27r^6)\right.
\nonumber\\
&\left.\qquad
+9(r^2-1)^3(3r^2+1)\log\left|\frac{r+1}{r-1}\right|\right].
\nonumber
\end{align}
for $a_{mn}$.

Next consider the corrections $B$. This term is already of 
the order $\mathcal{O}(\{\Plin(k)\}^2)$, and the non-vanishing 
contribution can be estimated without employing the perturbative 
calculations. Just applying 
the formulae (\ref{eq:integration_formula})
in Appendix \ref{appendix:formula} to Eq.~(\ref{eq:B_term}),  
we obtain
\begin{widetext}
\begin{eqnarray}
B(k,\mu)=\sum_{n=1}^4\,\, \sum_{a,b=1}^2 \mu^{2n}(-f)^{a+b} 
\frac{k^3}{(2\pi)^2}\int_0^\infty dr \int_{-1}^{+1} dx 
\,B^n_{ab}(r,x)\,\frac{P_{a2}\left(k\sqrt{1+r^2-2rx}\right)P_{b2}(kr)}
{(1+r^2-2rx)^a}, 
\label{eq:B_PT_formula}
\end{eqnarray}
\end{widetext}
where $P_{12}(k)=\Pdv(k)$ and $P_{22}(k)=\Pvv(k)$. 
The non-vanishing coefficients $B^n_{ab}$ are
\begin{align}
&B^1_{11}=\frac{r^2}{2}(x^2-1),
\nonumber
\end{align}
\begin{align}
&B^1_{12}=\frac{3r^2}{8}(x^2-1)^2,
\nonumber
\end{align}
\begin{align}
&B^1_{21}=\frac{3r^4}{8}(x^2-1)^2,
\nonumber
\end{align}
\begin{align}
&B^1_{22}=\frac{5r^4}{16}(x^2-1)^3,
\nonumber
\end{align}
\begin{align}
&B^2_{11}=\frac{r}{2}(r+2x-3rx^2),
\nonumber
\end{align}
\begin{align}
&B^2_{12}=-\frac{3r}{4}(x^2-1)(-r-2x+5rx^2),
\nonumber
\end{align}
\begin{align}
&B^2_{21}=\frac{3r^2}{4}(x^2-1)(-2+r^2+6rx-5r^2x^2),
\nonumber
\end{align}
\begin{align}
&B^2_{22}=-\frac{3r^2}{16}(x^2-1)^2(6-30rx-5r^2+35r^2x^2),
\nonumber
\end{align}
\begin{align}
&B^3_{12}=\frac{r}{8}\left\{4x(3-5x^2)+r(3-30x^2+35x^4)\right\},
\nonumber
\end{align}
\begin{align}
&B^3_{21}=\frac{r}{8}\left[-8x+r\left\{-12+36x^2+12rx(3-5x^2)\right.\right.
\nonumber\\
&\left.\left.\qquad\qquad\qquad\qquad\qquad
+r^2(3-30x^2+35x^4)\right\}\right],
\nonumber
\end{align}
\begin{align}
&B^3_{22}=\frac{3r}{16}(x^2-1)\left[-8x+r\left\{-12+60x^2
\right.\right.
\nonumber\\
&\left.\left.\qquad
+20rx(3-7x^2)+ 5r^2(1-14x^2+21x^4)\right\}\right],
\nonumber
\end{align}
\begin{align}
&B^4_{22}=\frac{r}{16}\left[8x(-3+5x^2)-6r(3-30x^2+35x^4)
\right.
\nonumber\\
&\quad\quad\quad+6r^2x(15-70x^2+63x^4)
\nonumber\\
&\quad\quad\quad\left.+r^3\left\{5-21x^2(5-15x^2+11x^4)\right\}\right].
\nonumber
\end{align}

The expression (\ref{eq:B_PT_formula}) is still non-perturbative 
in the sense that we do not perturbatively treat the power spectra 
$\Pdv$ and $\Pvv$ in the integrand. For the leading-order calculation,  
we simply apply the linear-theory calculation to these quantities, and 
replace both $\Pdv$ and $\Pvv$ with the linear spectrum $\Plin$.

\section{Some useful formulae for integrals}
\label{appendix:formula}

In this Appendix, we give the integral formulae 
used in Appendix \ref{appendix:PT_calc_correction}  
to derive the perturbative expressions for the correction $A$ and $B$.

Let us first consider the integral of an arbitrary 
scalar function $f(\bfk,\bfp)$ times some vectors over $\bfp$.  
A simple example of the integrand are   
$p_ip_j\,f(\bfk,\bfp)$, where subscript $i,\,j$ selects $x$-,\,$y$- or 
$z$-direction. The rotationally invariant properties of the integral 
implies that the resultant form of the integral is given by  
\begin{align}
&\int \frac{d^3\bfp}{(2\pi)^3}\,p_i p_j\, f(\bfk,\bfp) = 
P\,\delta_{ij} + Q\,k_ik_j, 
\label{eq:pi_pj}
\end{align}
irrespective of the functional form of $f(\bfk,\bfp)$. 
The coefficients $P$ and $Q$ are 
obtained by contracting the above integral with $\delta_{ij}$ and $k_ik_j$, 
and are the functions of $k=|\bfk|$.  We have 
\begin{align}
&P=\frac{k^5}{(2\pi)^2}\int dr\,r^2 \int_{-1}^1 dx x\,
\frac{r^2}{2}\left(1-x^2\right)\,f(k,r,x),
\nonumber\\
&Q=\frac{k^3}{(2\pi)^2}\int dr\,r^2 \int_{-1}^1 dx x\,
\frac{r^2}{2}\left(3x^2-1\right)\,f(k,r,x), 
\nonumber
\end{align}
where we write $p=kr$ and $\bfk\cdot\bfp=k^2r x$. Thus, as a special case 
with $i=j=z$, we get the following formula: 
\begin{align}
\int \frac{d^3\bfp}{(2\pi)^3}\,p_z^2\, f(\bfk,\bfp) =
P\,+ (k\mu)^2\,Q
\end{align}
with $k_z=k\,\mu$.

The above procedure can be generalized to the cases of integrals 
involving arbitrary numbers of multiplicative vectors. 
As a useful formula, we here explicitly write down the result 
summing up the integrals of arbitrary scalar functions $f_n$ times 
the power $p_z^n$ up to the sixth order:  
\begin{align}
&  \sum_{n=0}^{6}\,\int\frac{d^3\bfp}{(2\pi)^3}\,
p_z^n\,f_n(\bfk,\bfp)=
\frac{k^3}{(2\pi)^2}\sum_{m,n=0}^{6}\mu^n\,
\nonumber\\
&
\times \int_0^\infty dr\,r^2\int_{-1}^{+1} dx\,
(kr)^m\,G_{nm}(x)\,\,f_m(k,r,x),
\label{eq:integration_formula}
\end{align}
The non-vanishing coefficients $G_{nm}$ as functions of $k,\,r$ and $x$ 
are summarized as follows:
\begin{align}
&G_{00}=1,
\nonumber
\end{align}
\begin{align}
&G_{02}=-\frac{1}{2}(x^2-1),
\nonumber
\nonumber
\end{align}
\begin{align}
&G_{04}=\frac{3}{8}(x^2-1)^2, 
\nonumber
\nonumber
\end{align}
\begin{align}
&G_{06}=-\frac{5}{16}(x^2-1)^3,
\nonumber
\nonumber
\end{align}
\begin{align}
&G_{11}= x,
\nonumber
\nonumber
\end{align}
\begin{align}
&G_{13}=-\frac{3}{2}x(x^2-1),
\nonumber
\nonumber
\end{align}
\begin{align}
&G_{15}=\frac{15}{8}x(x^2-1)^2,
 \nonumber
\nonumber
\end{align}
\begin{align}
&G_{22}=\frac{1}{2}(3x^2-1),
\nonumber
\nonumber
\end{align}
\begin{align}
& G_{24}=-\frac{3}{4}(5x^4-6x^2+1),
\nonumber
\nonumber
\end{align}
\begin{align}
& G_{26}=\frac{15}{16}(7x^2-1)(x^2-1)^2,
 \nonumber
\nonumber
\end{align}
\begin{align}
&G_{33}=\frac{1}{2}x(5x^2-3),
\nonumber
\end{align}
\begin{align}
& G_{35}=-\frac{5}{4}x(7x^4-10x^2+3),
 \nonumber
\end{align}
\begin{align}
&G_{44}=\frac{1}{8}(35x^4-30x^2+3),
 \nonumber
\end{align}
\begin{align}
&G_{46}=\frac{15}{16}(-21x^6+35x^4-15x^2+1),
 \nonumber
\end{align}
\begin{align}
&G_{55}=\frac{1}{8}(63x^4-70x^2+15),
 \nonumber
\end{align}
\begin{align}
& G_{66}=\frac{1}{16}(231x^6-315x^4+105x^2-5).
\nonumber
\end{align}

\section{Covariance between multipole power spectra }
\label{appendix:covariance}

Here, we give the explicit expression for covariance between 
multipole power spectra used in the MCMC analysis in 
Sec.~\ref{subsec:recovery_DA_H_f}. 

Neglecting the non-Gaussian contribution, the non-vanishing part of 
the covariance only appears at the diagonal components (i.e., 
correlation between the same Fourier modes),  which are expressed as 
(e.g., Ref.~\cite{Yamamoto:2002bc,Yamamoto:2005dz})
\begin{align}
&\mbox{Cov}_{\ell,\ell'}(k)=\frac{2}{N_k}\frac{(2\ell+1)(2\ell'+1)}{2}
\nonumber\\
&\quad\quad\times
\int_{-1}^1d\mu \,\mathcal{P}_\ell(\mu)\mathcal{P}_{\ell'}(\mu)
\left\{P^{\rm(S)}(k,\mu)+\frac{1}{\overline{n}_{\rm g}}\right\}^2,
\label{eq:def_Cov}
\end{align}
where $N_k$ is the number of Fourier modes within a given bin at $k$, 
given by $N_k=4\pi\,k^2\Delta k/(2\pi/V_s^{1/3})^3$ with $\Delta k$ and 
$V_s$ being the bin width and survey volume, respectively.

For an analytic estimate of the covariance, we assume that 
the power spectrum is simply written as 
$P^{\rm(S)}(k,\mu)=(1+\beta\,\mu^2)^2b^2\,P_{\rm lin}(k)$, where $b$ is 
the linear bias parameter and $\beta$ is defined by $\beta\equiv f/b$. 
Substituting this into Eq.~(\ref{eq:def_Cov}), we obtain 
\begin{align}
&\mbox{Cov}_{0,0}(k)=\frac{2}{N_k}\,
\nonumber\\
&\times\left[
\left(1+\frac{4}{3}f+\frac{6}{5}\beta^2+\frac{4}{7}\beta^3+
\frac{1}{9}\beta^4\right)
\left\{b^2\,P_{\rm lin}(k)\right\}^2
\right.
\nonumber\\
&\qquad\qquad
\left.+\frac{2}{\overline{n}_{\rm g}}\left(1+\frac{2}{3}\beta
+\frac{1}{5}\beta^2\right)
b^2\,P_{\rm lin}(k)+\frac{1}{\overline{n}_{\rm g}^2}\right]
\end{align}
for $(\ell,\ell')=(0,0)$, 
\begin{align}
&\mbox{Cov}_{0,2}(k)=\frac{2}{N_k}\,
\nonumber\\
&\times\left[
\left(\frac{8}{3}\beta+\frac{24}{7}\beta^2+\frac{40}{21}\beta^3+
\frac{40}{99}\beta^4\right)\left\{b^2\,P_{\rm lin}(k)\right\}^2
\right.
\nonumber\\
&\qquad\qquad
\left.+\frac{2}{\overline{n}_{\rm g}}\left(\frac{4}{3}\beta
+\frac{4}{7}\beta^2\right)
b^2\,P_{\rm lin}(k)\right]
\end{align}
for $(\ell,\ell')=(0,2)$ or $(2,0)$, and 
\begin{align}
&\mbox{Cov}_{2,2}(k)=\frac{2}{N_k}\,
\nonumber\\
&\times\left[
\left(5+\frac{220}{21}\beta+\frac{90}{7}\beta^2+\frac{1700}{231}\beta^3+
+\frac{2075}{1287}\beta^4\right)
\right.
\nonumber\\
&\qquad\quad\times\left\{b^2\,P_{\rm lin}(k)\right\}^2
+\frac{2}{\overline{n}_{\rm g}}\left(5+\frac{220}{21}\beta
+\frac{30}{7}\beta^2\right)
\nonumber\\
&\qquad\quad
\left.\times 
b^2\,P_{\rm lin}(k)+\frac{5}{\overline{n}_{\rm g}^2}\right]
\end{align}
for $(\ell,\ell')=(2,2)$.


\end{document}